\documentclass[aps, prd, showpacs, floatfix, superscriptaddress, twocolumn, nofootinbib, preprintnumbers, longbibliography]{revtex4-2}
\usepackage{orcidlink}
\usepackage{lipsum, multirow, microtype, amsmath, amssymb, newfloat, bm, color}
\usepackage{graphicx}
\usepackage{natbib}
\usepackage{comment}
\usepackage{hyperref}
\usepackage{hypcap, mathrsfs, placeins, etoolbox}
\usepackage{dcolumn}
\usepackage[dvipsnames]{xcolor}
\usepackage[utf8]{inputenc}
\usepackage[scaled = 0.92]{helvet}
\usepackage{inconsolata}
\usepackage{enumitem}
\usepackage{xr}
\usepackage{cleveref}
\usepackage{booktabs}
\setlist[itemize]{leftmargin = 9pt, labelsep = 0.31em, itemsep = 0.055em}
\graphicspath{{Figures/}} 
\hypersetup{
colorlinks = true,
citecolor = cyan,
linkcolor = magenta,
urlcolor = NavyBlue,
allbordercolors = {0 0 0},
pdfborderstyle = {/S/U/W 1}
}
\def \Hz	    {\mathrm{Hz}}

\def \Mc        {\mathcal{M}_C}
\def \flow	    {f_{\mathrm{low}}}

\def \MM        {\mathcal{MM}}

\newcommand{\IAR}{Institute of Advanced Research, Koba Institutional Area, Gandhinagar - 382 426, India.\vspace*{0.125cm}}
\newcommand{\SXC}{St.Xavier's College (Autonomous), Navrangpura, Ahmedabad - 380 009, India.\vspace*{0.125cm}}

\newcommand{\QWT}{Centre for Development of Telematics (C-DOT), Electronics City Phase 1, Electronic City, Bengaluru, Karnataka 560100, India.
\vspace*{0.125cm}}
\newcommand{\NIKHEF}{Nikhef, Science Park 105, 1098 XG Amsterdam, The Netherlands.}

\newcommand{\IWF}{Space Research Institute, Austrian Academy of Sciences, Schmiedlstrasse 6, 8042 Graz, Austria}
\newcommand{\UMD}{Department of Physics, University of Massachusetts, Dartmouth, MA 02747, USA.}

\begin{document}
\title{Neural Network–Guided Parameter-Space Constraints for Gravitational-Wave Searches from Binary Black Holes}
\author{Chetan Verma} 
\email{chetan.verma@sxca.edu.in} \affiliation{\IAR}
\affiliation{\SXC} 
\author{Amit Reza {\orcidlink{0000-0001-7934-0259}}
} 
\email{amit.reza@oeaw.ac.at} \affiliation{\IWF}
\affiliation{\NIKHEF} 
\author{Gurudatt Gaur \vspace*{0.15cm}} 
\email{gurudatt.gaur@sxca.edu.in} \affiliation{\SXC} 
%
\author{Dilip Krishnaswamy \vspace*{0.15cm}} 
\email{dilip@ieee.org} \affiliation{\QWT}
\author{Sarah Caudill \vspace*{0.15cm}} 
\email{scaudill@umassd.edu}
\affiliation{\UMD}
\begin{abstract}
The detection of gravitational waves (GWs) from compact binary coalescences (CBCs) using matched filtering remains computationally demanding because detector data must be correlated with a large number of analytical signals (called template waveforms) spanning a high-dimensional intrinsic parameter space. In our previous work~\cite{Verma2022}, we demonstrated that a convolutional neural network (CNN) not only classifies noisy signals against pure noise but can also tightly constrain the intrinsic parameter space—such as the mass regime (defined as a patch) of a true, non-spinning binary black hole (BBH) signal, enabling a matched-filter search region that is narrower and reduces computational cost. In this work, we extend the framework to aligned-spin BBH systems and systematically investigate how different parameter-space representations influence CNN-based patch identification. Using waveforms generated with the \texttt{IMRPhenomD} model over the aligned-spin BBH parameter space and assuming the Advanced LIGO design sensitivity, we first show that a CNN trained solely on the aligned-spin template bank achieves more than $99.8\%$ signal--noise classification accuracy on independently generated uniformly sampled BBH signals, demonstrating that additional uniformly sampled training data are unnecessary for this task. We then partition the common aligned-spin template bank into four approximately balanced patches using a Principal Component Analysis (PCA)-based quantile scheme and evaluate five physically motivated parameter-space representations for CNN-based patch identification. Among the parameterizations investigated, the chirp mass--duration $(\mathcal{M}_c,\tau)$ representation achieves the highest average patch-identification accuracy (93.2\%), followed by chirp mass (91.6\%) and component masses (90.5\%). In contrast, the post-Newtonian coordinate systems $(\tau_0,\tau_3)$ and $(\theta_0,\theta_3,\theta_{3s})$ yield substantially lower accuracies. These results demonstrate that the existing template bank is sufficient for training and detecting the noisy signal with high accuracy. However, the choice of the intrinsic parameter space is critical for effectively constraining the intrinsic parameters of the true signal. 
\end{abstract}

\maketitle

\newpage
\setcounter{page}{1}


\section{Introduction}
Gravitational-wave (GW) astronomy has emerged as a fundamentally new observational window onto the Universe, enabling the direct study of compact-object dynamics and strong-field gravity. Following the first detection of a binary black hole (BBH) merger, GW150914, by the Advanced Laser Interferometer Gravitational-Wave Observatory (LIGO) in 2015~\cite{abbott2016gw150914}, the field has rapidly matured into a precision observational science. The global network of ground-based interferometric detectors consisting of Advanced LIGO, Virgo, and KAGRA has now reported more than one hundred compact binary coalescence (CBC) events~\cite{gwtc1,gwtc3}, including BBH, binary neutron star (BNS), and neutron star--black hole (NSBH) mergers. These observations have enabled stringent tests of general relativity in the highly dynamical strong-field regime~\cite{TestGR2021}, improved our understanding of compact-object populations and formation channels~\cite{AstroImplications}, and initiated the era of multimessenger astronomy through the binary neutron star event GW170817~\cite{abbott2017gw170817}. 

Ground-based GW observatories are sensitive primarily in the frequency band $\sim 10$--$10^3$ Hz, where the dominant sources are coalescing compact binaries. Detection in this regime is presently carried out using matched-filtering-based search pipelines such as GstLAL~\cite{Cody}, PyCBC~\cite{usman2016pycbc}, MBTA~\cite{MBTA}, and SPIIR~\cite{SPIIR}. These pipelines correlate detector strain data against large banks of theoretically modeled waveform templates spanning the intrinsic parameter space of the source. For CBC signals, this parameter space includes, at a minimum, the component masses and spin degrees of freedom of the binary. Current matched-filter searches for compact binary coalescences, including both low-latency and offline analyses performed by pipelines such as GstLAL and PyCBC, predominantly employ aligned-spin waveform template banks~\cite{Mukherjee2021,usman2016pycbc}. Although the gravitational waveform formally depends on the two component spins $(\chi_1,\chi_2)$, the dominant spin effects in aligned-spin systems can be effectively captured by the mass-weighted effective spin parameter
\begin{equation}
\chi_{\mathrm{eff}} = \frac{m_1 \chi_1 + m_2 \chi_2}{m_1 + m_2}.
\end{equation}
Consequently, template placement for aligned-spin searches is commonly performed in an effective three-dimensional intrinsic parameter space spanned by $(m_1,m_2,\chi_{\mathrm{eff}})$ or equivalent reduced-spin coordinates~\cite{Emily2013}.

Matched filtering remains the optimal detection strategy for signals of known morphology embedded in stationary Gaussian noise~\cite{helstrom1994elements, Findchirp}. However, its computational cost scales directly with the number of templates required to densely cover the parameter space at a specified minimal match. For aligned-spin BBH searches with Advanced LIGO sensitivity and a typical minimal match of $\MM \sim 0.97$, the number of templates can exceed $10^6$~\cite{Harry2009, Satya}. The computational burden becomes even more severe because the filtering cost per template scales approximately as $N_s \log N_s$, where $N_s = f_s T$ is the number of waveform samples, determined by the sampling frequency $f_s$ and signal duration $T$. In current CBC searches, BBH and BNS systems are generally analyzed using distinct template banks and search configurations tailored to their respective waveform durations and frequency content~\cite{usman2016pycbc, Cody}. While the longer duration of BNS signals can increase filtering cost, this is partially mitigated through lower sampling rates and multiband search strategies. However, the overall computational burden associated with increasingly large template banks and higher-dimensional waveform models remains a central challenge for present and future GW searches.

This challenge will become substantially more acute in the era of next-generation GW observatories such as the Einstein Telescope (ET)~\cite{Punturo2010, Maggiore2020} and Cosmic Explorer (CE)~\cite{Reitze2019}. These detectors are expected to achieve order-of-magnitude improvements in strain sensitivity and extend the observational band to frequencies as low as $\sim 1$--$3$ Hz. Lower-frequency sensitivity dramatically increases waveform durations in the band, particularly for low-mass systems, thereby increasing the size of the data segments required for filtering. Simultaneously, future searches will require increasingly sophisticated waveform models incorporating higher-order harmonics, spin-induced precession, eccentricity, tidal effects, and beyond-leading-order relativistic corrections~\cite{Husa2016}. The incorporation of increasingly realistic physical effects --- including spin precession, higher-order modes, eccentricity, and tidal interactions --- leads to substantially more complex waveform manifolds~\cite{Ajith2011, Husa2016}. Although aligned-spin searches admit relatively efficient reduced-spin parameterizations, robust template-placement strategies for generic higher-dimensional waveform families remain incompletely understood. As a result, current approaches frequently rely on stochastic template placement and computationally expensive sampling methods. The associated increase in computational cost presents a major challenge for sustaining low-latency GW searches and rapid multimessenger follow-up in the era of next-generation detectors.

Recent years have therefore witnessed growing interest in applying machine learning and deep learning methods to GW data analysis~\cite{Cuoco2021}. Convolutional neural networks (CNNs), recurrent neural networks (RNNs), autoencoders, and transformer-based architectures have all been explored for GW detection, denoising, glitch classification, and parameter estimation~\cite{gabbard2018matching, krastev2020real, George:2016hay, George:2017pmj, gebhard2019convolutional, wei2020gravitational, wang2020gravitational}. Among the pioneering works, George and Huerta~\cite{George:2016hay, George:2017pmj} introduced the concept of ``deep filtering,'' demonstrating that neural networks can detect GW signals in real time with sensitivity comparable to matched filtering in Gaussian noise. Gabbard \textit{et al.}~\cite{gabbard2018matching} further showed that CNNs can reproduce matched-filter sensitivities for BBH detection, while Krastev~\cite{krastev2020real} extended such approaches to real-time BNS detection.

Despite these advances, most deep-learning-based approaches formulate GW searches as binary classification or regression problems. While such methods can rapidly determine whether a signal is present, they generally do not provide a mechanism to reduce the size of the matched-filter search space. Consequently, they are not readily integrated into existing matched-filtering pipelines as computational accelerators. Furthermore, fully replacing matched filtering with deep learning remains difficult because of the stringent robustness requirements imposed by real detector noise, non-stationarity, and the need for accurate parameter estimation.

In our previous work~\cite{Verma2022}, we proposed a hierarchical framework designed specifically to address this limitation. Instead of replacing matched filtering, we employed deep learning as a pre-selection tool to constrain the intrinsic parameters of true GW signals within coarse regions (defined as ``patches''). The hierarchical framework consisted of two stages: a first-stage CNN that distinguished GW signals from pure noise, and a second-stage CNN that identified the parameter-space patch to which the true signal belonged. For non-spinning BBH signals in the mass range $[10,70]\, M_{\odot}$, we demonstrated $\sim 99\%$ accuracy in signal-versus-noise classification and $\gtrsim 97\%$ average accuracy in patch identification. This framework could be used to restrict the subsequent matched-filter search to templates within the identified patch, thereby substantially reducing the total number of matched-filter operations. Moreover, the patch boundaries help establish an accurate prior for subsequent Bayesian parameter estimation. 

The zero-spin approximation used in that study ~\cite{Verma2022} represents a significant physical limitation. Astrophysical black holes are expected in general to possess non-negligible spin angular momentum, and spin effects can substantially alter the morphology of GW signals through spin-orbit coupling and orbital hang-up effects~\cite{Ajith2011, Damour2016}. Aligned spins modify both the phase evolution and duration of the waveform, thereby affecting the structure of the template bank itself. Extending the patch-identification framework to spinning systems is therefore essential for realistic deployment in production GW search pipelines.

In this work, we present a comprehensive extension of the CNN-based patch-identification framework to aligned-spin BBH systems. We consider the four-dimensional intrinsic parameter space defined by
\[
m_1,m_2 \in [5,50]\,M_\odot,
\qquad
\chi_1,\chi_2 \in [-0.99,0.99],
\]
using the Advanced LIGO design sensitivity PSD. We systematically investigate patch-construction strategies in one-, two-, and three-dimensional parameter spaces using several physically motivated coordinate systems, including chirp mass $(\Mc)$, signal duration $(\tau)$, mass ratio $(q)$, the post-Newtonian timing coordinates $(\tau_0,\tau_3)$, and the three-dimensional spin-inclusive coordinate system $(\theta_0,\theta_3,\theta_{3s})$. Using a common aligned-spin template bank, we construct parameter-space patches in several one-, two-, and three-dimensional coordinate representations and systematically compare their CNN classification performance. This enables us to assess how the choice of parameter-space representation influences waveform separability and the effectiveness of hierarchical machine-learning-assisted gravitational-wave searches.

Our results demonstrate that the effectiveness of deep-learning-assisted patch identification depends strongly on the choice of parameter-space representation. In particular, we find that parameterizations aligned with dominant waveform morphology and parameter-space degeneracy directions yield significantly improved classification performance. Furthermore, we show that the proposed framework can substantially reduce the effective matched-filtering cost while maintaining high patch-identification accuracy, thereby providing a scalable pathway toward low-latency GW searches in the era of next-generation detectors and increasingly sophisticated waveform models.

The remainder of this paper is organized as follows. Section~\ref{sec:theory} reviews the theoretical background, including matched filtering and the coordinate systems employed for patch construction. Section~\ref{sec:method} describes the CNN architecture, waveform generation procedure, and training methodology. Section~\ref{sec:results} presents the performance of the various patch schemes and analyzes the resulting classification accuracies. Section~\ref{sec:discussion} discusses the implications of our findings, limitations of the current framework, and prospects for future extensions. Finally, Section~\ref{sec:conclusion} summarizes our conclusions.
\section{Theoretical Background}
\label{sec:theory}

\subsection{Matched Filtering and the Template Bank}
\label{subsec:mf}

Matched filtering remains the optimal detection statistic for a GW signal of known form embedded in stationary Gaussian noise~\cite{helstrom1968statistical}. The inner product between the detector output $s(t)$ and a template waveform $h(t)$ is defined as:
\begin{equation}
  \langle s(t),\, h(t) \rangle = 4\,\mathrm{Re}
  \int_0^{\infty} \frac{\tilde{s}(f)\,\tilde{h}^*(f)}{S_n(f)}\,df \, ,
  \label{eq:inner_product}
\end{equation}
where $S_n(f)$ is the one-sided noise power spectral density (PSD), and tildes denote Fourier transforms. The matched-filter SNR is then:
\begin{equation}
  \rho(t) = \frac{\langle s(t),\, h(t) \rangle}{\sigma} \, ,
  \label{eq:snr}
\end{equation}
where $\sigma^2 = \langle h(t),\, h(t) \rangle$ is the template norm. A trigger is generated whenever $\rho(t)$ exceeds a chosen threshold $\rho_{th}$. The optimal SNR for a given template is:
\begin{equation}
  \rho_{opt} = \sqrt{4 \int_{f_{\min}}^{f_{\mathrm{high}}}
    \frac{|\tilde{h}(f)|^2}{S_n(f)}\,df}.
  \label{eq:ropt}
\end{equation}

To ensure sensitivity to all signals within the target parameter space, the template bank is constructed such that the match—the normalized inner product between any signal and its nearest template—is at least equal to the minimal match ($\MM$). In current LVK aligned-spin compact binary coalescence (CBC) searches, template banks are typically constructed with a minimal match of $\MM \simeq 0.97$ and a low-frequency cutoff around $f_{low} \sim 20 \Hz$. For BBH searches over the relevant mass and spin ranges, the resulting stochastic template banks can contain $\mathcal{O}(10^6)$ templates.

\subsection{Intrinsic Parameter Space and Coordinate Systems}
\label{subsec:coords}

The waveform of an aligned-spin (non-precessing) BBH is determined by four intrinsic parameters: the component masses $m_1 \geq m_2$ and the dimensionless aligned-spin components $\chi_1, \chi_2 \in [-1,\,1]$. Several coordinate systems provide physically meaningful parameterizations of this space, and some offer substantially reduced correlations between parameters---a property that is valuable for defining compact, well-shaped patches.

\medskip\noindent\textbf{Component Mass Space ($m_1$-$m_2$):}
The most physically transparent parameterization. The physical constraint $m_1 \geq m_2 > 0$ restricts the accessible region to a triangular domain in the $m_1$-$m_2$ plane. In this space, uniform mass distributions yield a triangular geometry with a well-defined boundary, making patch construction geometrically natural.

\medskip\noindent\textbf{Chirp Mass and Related Quantities:}
The chirp mass, defined as:
\begin{equation}
  \Mc = \frac{(m_1 m_2)^{3/5}}{(m_1 + m_2)^{1/5}},
  \label{eq:chirp_mass}
\end{equation}
is the combination of masses that most directly controls the frequency evolution of the waveform at leading Post-Newtonian (PN) order. The mass ratio $q = m_1/m_2 \geq 1$ and the total mass $M = m_1 + m_2$ are complementary parameters. Signal duration $\tau$ is strongly correlated with $\Mc$, providing a natural second axis for 2D patch construction.

\medskip\noindent\textbf{Post-Newtonian Timing Parameters ($\tau_0$-$\tau_3$):}
The post-Newtonian chirp-time coordinates $(\tau_0,\tau_3)$ were introduced as convenient parameters for template-bank construction in compact-binary searches~\cite{Satya}. At the leading PN order, they are defined as
\begin{align}
\tau_0 &= \frac{5}{256\,\eta\,(\pi f_0)^{8/3} M^{5/3}},
\\[4pt]
\tau_3 &= \frac{\pi}{8\,\eta\,(\pi f_0)^{5/3} M^{2/3}},
\end{align}
where $M = m_{1} + m_{2}$ is the total mass, $\eta = m_{1} m_{2} / M^{2}$ is the symmetric mass ratio, and $f_{0}$ is a reference frequency, typically chosen as the low-frequency cutoff $\flow$.

\medskip\noindent\textbf{Three-Dimensional PN Spin Coordinates ($\theta_0$-$\theta_3$-$\theta_{3s}$):}
For aligned-spin systems, the intrinsic parameter space can be extended using spin-dependent PN phase coordinates such as $(\theta_0,\theta_3,\theta_{3s})$~\cite{Ajith2014}. These coordinates are constructed from combinations of the post-Newtonian phase coefficients appearing in the frequency-domain inspiral waveform and are designed to reduce variations in the local template-space metric across the bank. In particular, the coordinate $\theta_{3s}$ captures the dominant aligned-spin contribution through an effective-spin combination closely related to $\chi_{\mathrm{eff}}$. The $(\theta_0,\,\theta_3,\,\theta_{3s})$ representation approximately separates mass-dominated and spin-dominated waveform variations, leading to a more uniform distribution of templates in the transformed parameter space. Consequently, patch construction in this coordinate system can reduce class imbalance and classification ambiguity near patch boundaries.

\subsection{Patch Construction Philosophy}
\label{subsec:patch_philosophy}

Patch construction partitions the intrinsic parameter space into non-overlapping regions (patches), each of which defines an independent target class for the second-stage CNN classifier. From the perspective of matched filtering, an effective patching scheme should group together waveforms with similar morphology while maintaining clear separability between neighboring patches. The principal considerations governing patch construction are:

\begin{enumerate}[leftmargin=*,label=(\arabic*)]

\item \textbf{Boundary-induced ambiguity:}
Classification errors occur predominantly near patch boundaries, where neighboring regions contain waveforms with small mismatches, resulting in highly similar phase evolution and time-frequency morphology. Patch boundaries aligned approximately along directions of weak waveform variation can reduce inter-patch degeneracy and improve classification performance.

\item \textbf{Template-density uniformity:}
Large variations in template density across patches can lead to class imbalance during CNN training, potentially biasing the classifier toward densely populated regions of the parameter space. Coordinate systems that reduce variations in the local template-space metric, such as $(\tau_0,\tau_3)$ and $(\theta_0,\theta_3,\theta_{3s})$, generally yield more uniform template distributions and better-balanced training sets.

\item \textbf{Geometric regularity:}
Patches with simple, approximately regular geometries in the chosen coordinate system facilitate more stable partitioning of the parameter space and reduce the complexity of the resulting classification boundaries.

\item \textbf{Physical interpretability:}
Coordinate systems associated with physically meaningful waveform properties---for example, chirp-mass intervals corresponding to distinct inspiral timescales and frequency-evolution rates---allow a more direct interpretation of the resulting patches and facilitate straightforward mapping to matched-filter sub-banks in existing CBC search pipelines such as PyCBC and GstLAL or SGNL

\end{enumerate}

\section{Methodology}
\label{sec:method}

\subsection{Data Generation}
\label{subsec:data}
An aligned-spin template bank is generated using the \texttt{IMRPhenomD} waveform model with a minimal match of $\MM = 0.97$ and a low-frequency cutoff of $\flow = 20~\mathrm{Hz}$. The template bank spans the intrinsic parameter space described above and contains 90, 166 templates. For each template, the corresponding derived quantities, including chirp mass $\Mc$, waveform duration $\tau$, post-Newtonian chirp-time coordinates $(\tau_{0}, \tau_{3})$, and the transformed spin coordinates $(\theta_{0}, \theta_{3}, \theta_{3s})$, are subsequently computed. These template-bank points provide the common underlying parameter distribution for all patching schemes investigated in this work (see Sec.~\ref{subsec:patch_schemes}), enabling a consistent comparison of different parameter-space representations.

Since the present work focuses on patch identification within the intrinsic parameter space, all extrinsic parameters are held fixed throughout the dataset generation process. In particular, the sky location, polarization angle, orbital phase, inclination angle, and source distance are kept constant for all simulated signals. This is consistent with the construction of matched-filter template banks, which are defined only in terms of the intrinsic parameters of the waveform model.

For both the first-stage binary classification (signal versus noise) and the second-stage patch-identification task, waveform amplitudes are rescaled to obtain target optimal matched-filter signal-to-noise ratios (SNRs) in the range $\rho_{\mathrm{opt}}\in[10,20]$. Since the extrinsic parameters are fixed, the variation in SNR is introduced solely through amplitude rescaling rather than through changes in the physical source distance or detector orientation. For each simulated signal, the target optimal SNR is uniformly sampled from the specified interval, thereby exposing the CNN to a range of signal strengths representative of different detection conditions.

After partitioning the template bank into four patches using the PCA--quantile procedure described in Sec .~\ref {subsec:pca}, each patch contains approximately one quarter of the total templates. To construct balanced datasets while maintaining computational tractability, 12,000 templates are uniformly sampled from each patch for training, yielding a total of $4.8 \times 10^{4}$ training waveforms per parameter-space representation. An independent test set is generated by randomly selecting 2,000 templates from each patch, yielding a total of 8,000 test waveforms.

\subsection{PCA--Quantile Patch Construction}
\label{subsec:pca}
The parameter-space patches are constructed using Principal Component Analysis (PCA) \cite{pca_book}, followed by quantile-based partitioning. The objective is to generate approximately balanced patches while aligning the partition boundaries with the dominant direction of variation in the underlying parameter distribution.

For a set of parameter-space points represented by the feature vector $\mathbf{x} = (x_{1}, x_{2}, \ldots, x_{n}),$ the features are first standardized to have zero mean and unit variance. PCA is then applied to determine the direction of maximum variance in the parameter space. Each standardized point is projected onto the first principal component, $ z = \mathbf{w}^{\mathrm T}\mathbf{x},$ where $\mathbf{w}$ denotes the eigenvector corresponding to the largest eigenvalue of the covariance matrix. The resulting one-dimensional coordinate $z$ is partitioned into four equal-population intervals using the 25th, 50th, and 75th percentiles of its distribution. Each interval defines one patch class. Consequently, the patch boundaries correspond to constant values of the first principal component and appear as parallel straight lines (or hyperplanes in higher-dimensional spaces) in the original parameter space, unlike clustering algorithms such as $k$-means ~\cite{kmeans}  or Gaussian Mixture Models ~\cite{gmmbook}, which frequently produce highly unequal cluster populations for the parameter distributions considered in this work, the PCA--quantile approach guarantees approximately balanced patches by construction. This balance is particularly advantageous for CNN training, since it avoids class imbalance and enables a fair comparison between different parameter-space representations.

\subsection{Patch Schemes}
\label{subsec:patch_schemes}
We investigate five different parameter-space representations for our patch identification problem, spanning one-, two-, and three-dimensional coordinate systems. For each coordinate system, the sampled parameter-space points are partitioned into four approximately balanced patches using the PCA--quantile procedure described in the previous subsection ~\ref{subsec:pca}. The objective is to assess how the choice of parameterization influences waveform separability and, consequently, the CNN classification performance.

\begin{figure}
\centering
\includegraphics[width=\linewidth]{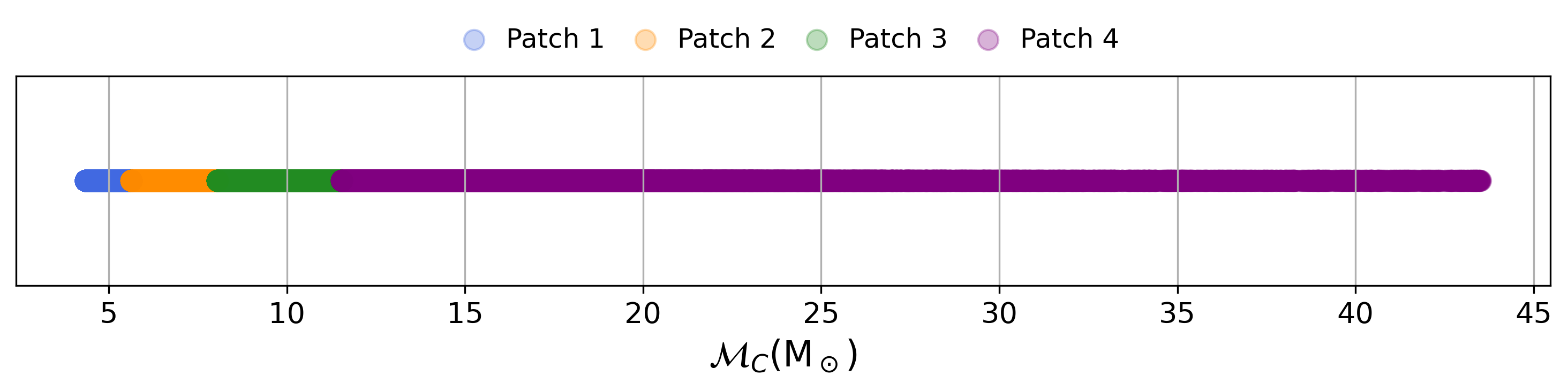}
\caption{Four-patch partition of the one-dimensional chirp-mass parameter space obtained using the PCA--quantile procedure.}
\label{fig:mc_patch}
\end{figure}

\medskip
\noindent\textbf{1D Patches ($\Mc$ Space):}
We first consider partitioning based solely on the chirp mass, $\Mc$. Since $\Mc$ governs the leading-order inspiral phase evolution and frequency sweep of compact-binary waveforms~\cite{Cutler1994}, it provides a natural one-dimensional representation of the dominant waveform morphology. Figure~\ref{fig:mc_patch} shows the resulting four-patch partition in chirp-mass space together with its projection onto the physical component-mass plane.

\medskip
\noindent\textbf{2D Patches:}

\begin{figure}
\centering
\includegraphics[width=\linewidth]{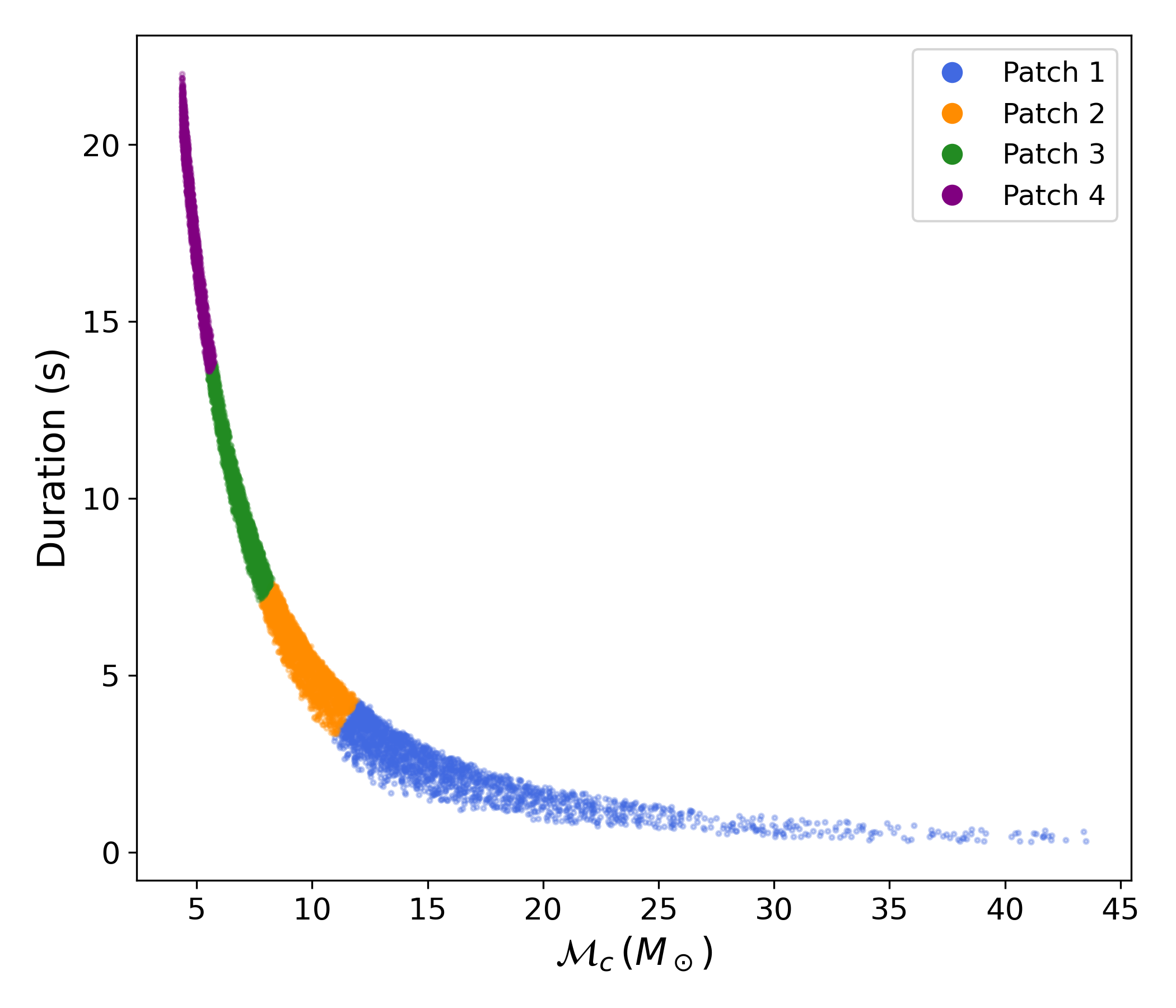}
\caption{Four-patch partition in the $(\Mc, \tau)$ parameter space obtained using the PCA--quantile procedure.}
\label{fig:mcdur_patch}
\end{figure}

\smallskip
\noindent\textit{$(\Mc,\tau)$ Space:}
The first two-dimensional representation employs chirp mass and waveform duration, where $\tau$ denotes the inspiral duration from the lower cutoff frequency $\flow$ to coalescence. Since waveform duration depends on both the component masses and aligned-spin effects, this representation captures additional waveform information beyond chirp mass alone. Figure~\ref{fig:mcdur_patch} illustrates the corresponding four-patch partition in the $(\Mc, \tau)$ parameter space obtained using the PCA--quantile procedure. 

\smallskip
\noindent\textit{$(m_{1}, m_{2})$ Space:}

\begin{figure}
\centering
\includegraphics[width=\linewidth]{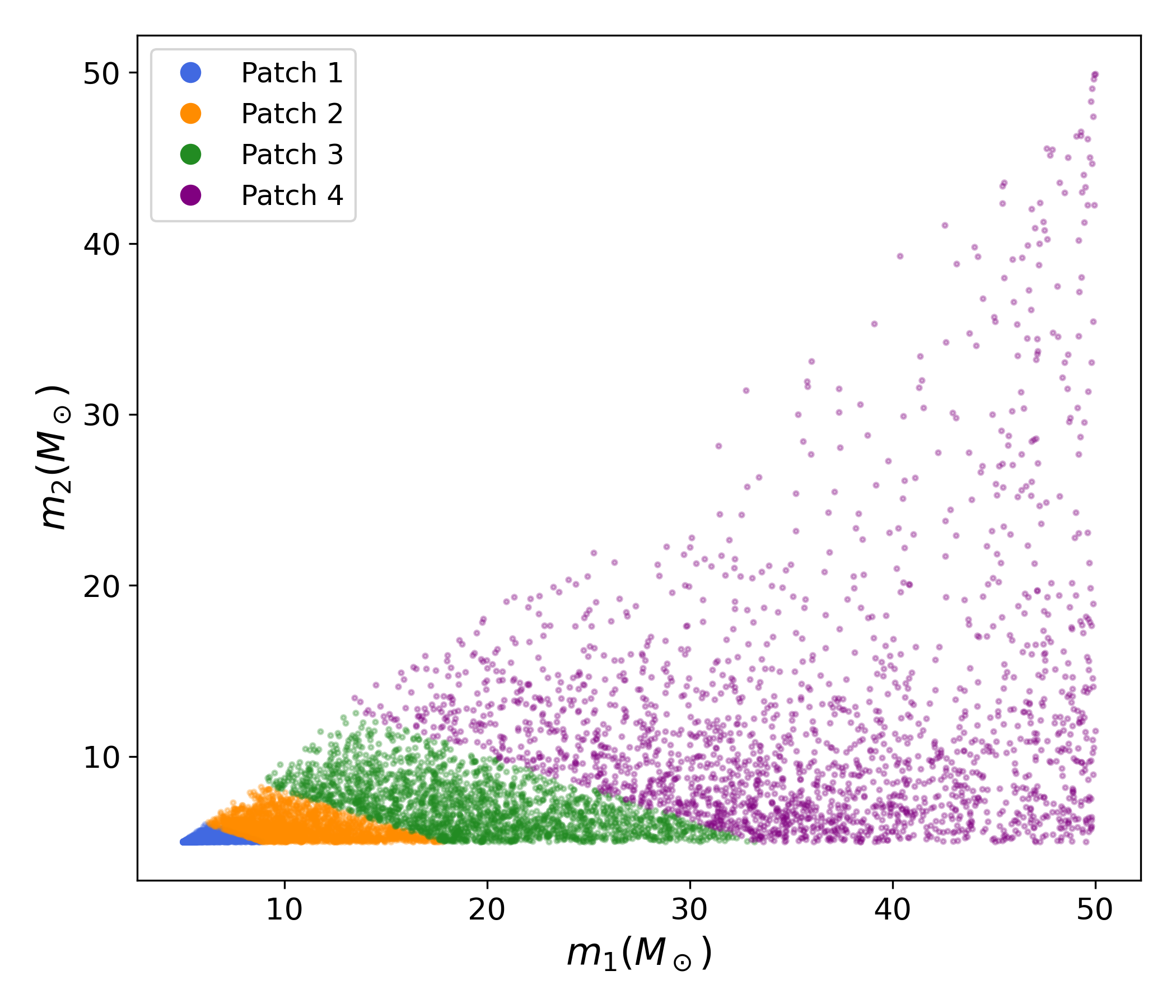}
\caption{Four-patch partition in the component-mass parameter space $(m_1,m_2)$.}
\label{fig:m1m2_patch}
\end{figure}

The second representation considers direct partitioning in the physical component-mass plane. The allowed parameter region is bounded by the constraint $m_{1} \ge m_{2}$, producing the familiar triangular domain shown in Fig.~\ref{fig:m1m2_patch}.

\smallskip
\noindent\textit{$(\tau_{0}, \tau_{3})$ Space:}

\begin{figure}
\centering
\includegraphics[width=\linewidth]{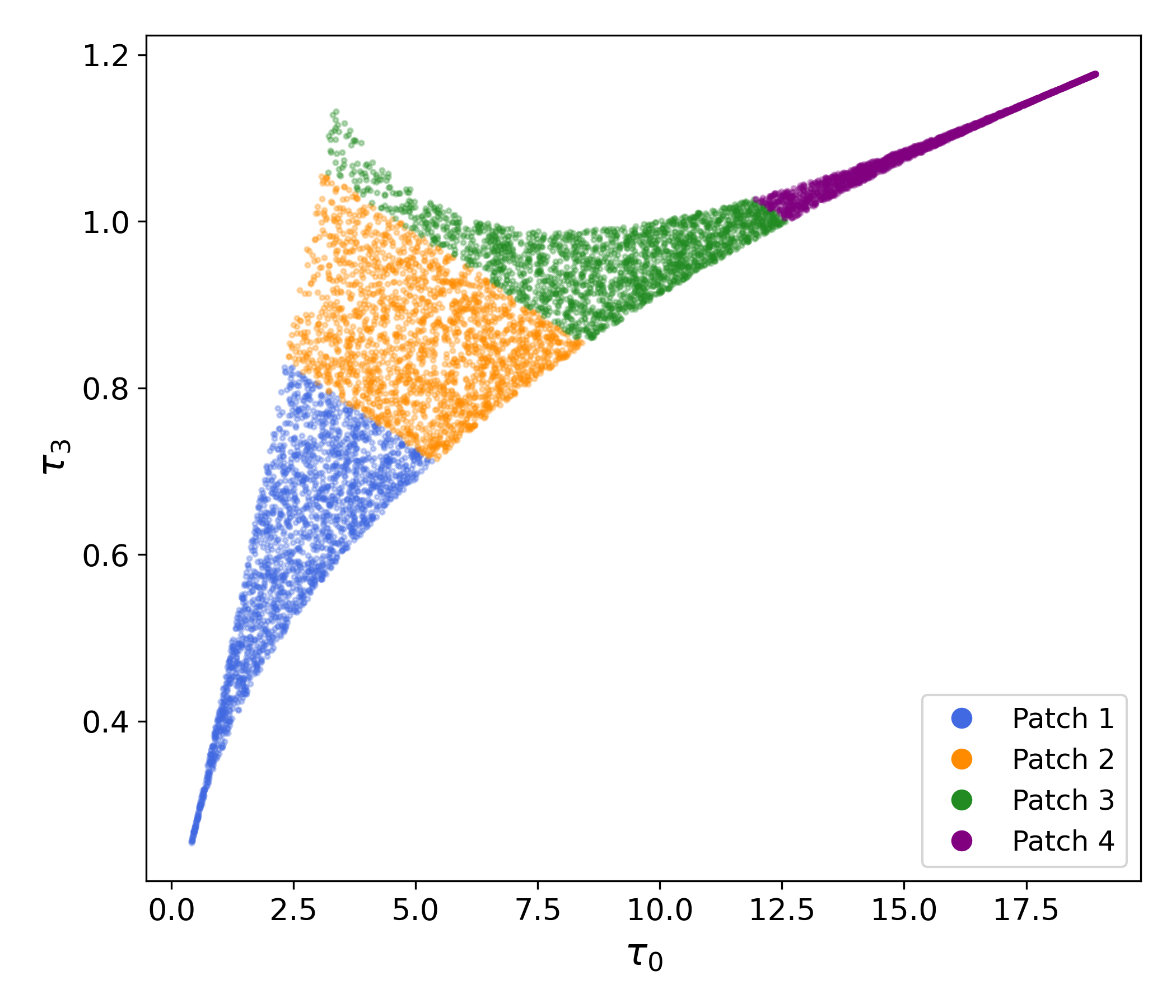}
\caption{Four-patch partition in the post-Newtonian chirp-time parameter space $(\tau_0,\tau_3)$.}
\label{fig:tau_patch}
\end{figure}

We also investigate the post-Newtonian chirp-time coordinates $(\tau_{0}, \tau_{3})$, which have been widely employed for template-bank construction because they provide an approximately flat parameter-space metric~\cite{Satya}. Figure~\ref{fig:tau_patch} shows the corresponding four-patch partition.

\medskip
\noindent\textbf{3D Patches ($\theta_{0}, \theta_{3}, \theta_{3s}$ Space):}

\begin{figure*}
\centering
\includegraphics[width=\linewidth]{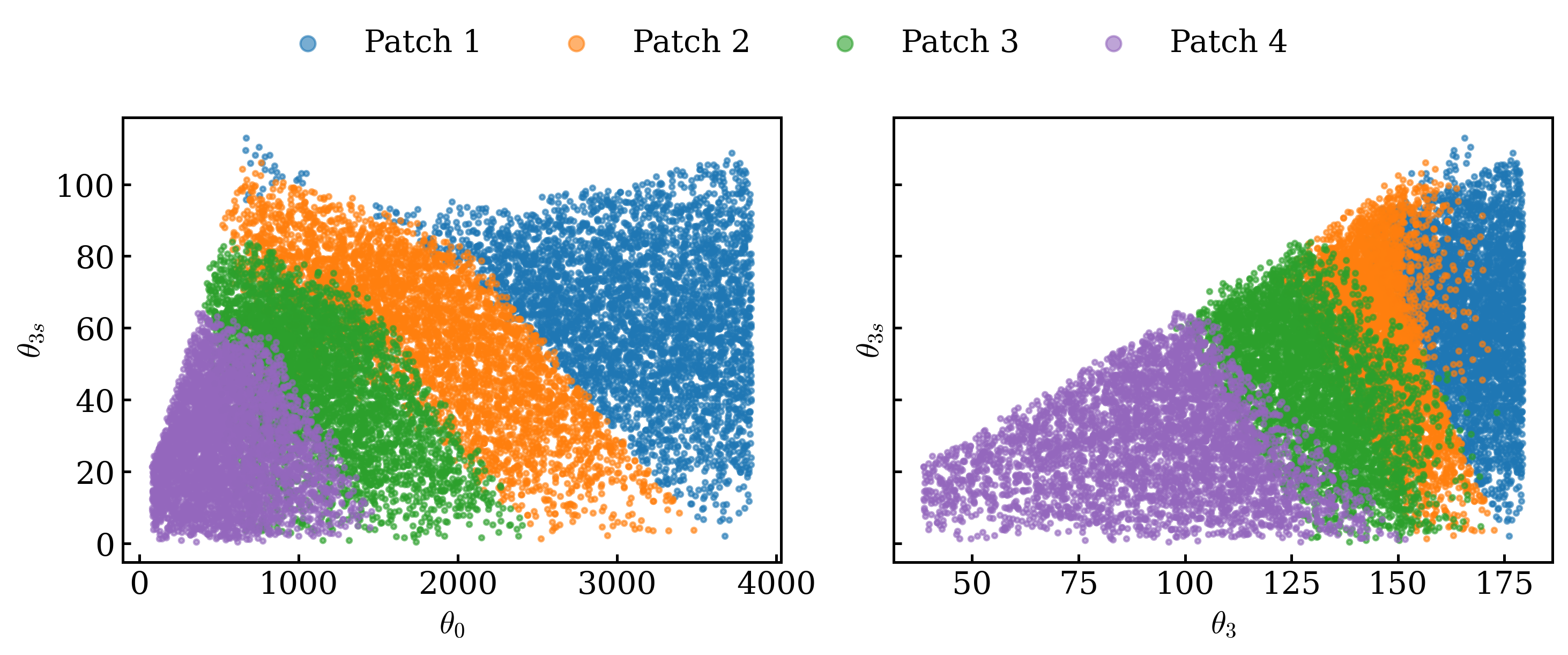}
\caption{Two-dimensional projections of the four-patch partition in the transformed parameter space $(\theta_{0}, \theta_{3}, \theta_{3s})$. The left panel shows the $(\theta_{0},\theta_{3s})$ projection, while the right panel shows the corresponding $(\theta_{3},\theta_{3s})$ projection.}
\label{fig:theta_patch}
\end{figure*}

Finally, we investigate a three-dimensional representation that explicitly incorporates aligned-spin effects through the transformed coordinates $(\theta_{0}, \theta_{3}, \theta_{3s})$~\cite{Ajith2014}. These coordinates are related to the post-Newtonian chirp-time parameters through
\begin{equation}
\theta_{0} = 2\pi f_{0} \tau_{0},\qquad
\theta_{3} = -2\pi f_{0} \tau_{3},\qquad
\theta_{3s}= 2\pi f_{0} \tau_{3S},
\end{equation}
where $f_{0}$ is the lower cutoff frequency. The spin-dependent coordinate $\theta_{3s}$ is computed from the reduced-spin parameter according to
\begin{equation}
\chi = \frac{48\pi \, \theta_{3s}}{113\, \theta_{3}},
\end{equation}
following Ref.~\cite{Ajith2014}. Figure~\ref{fig:theta_patch} presents two-dimensional projections of the resulting four-patch partition in the transformed parameter space.
The five parameterizations investigated above span physically motivated descriptions of the aligned-spin BBH intrinsic parameter space, ranging from directly observable waveform quantities to coordinates designed for template-bank construction. Comparing their CNN classification performance enables us to assess how the underlying parameter-space representation influences waveform separability in NN-guided gravitational-wave searches.

\subsection{CNN Architecture}
\label{subsec:cnn}
We employ the two-dimensional convolutional neural network (2D CNN) architecture introduced in our previous works~\cite{Verma2022,verma2024detection}. In those studies, 2D representations of the strain data achieved classification performance comparable to analogous 1D CNN architectures while requiring a relatively modest number of trainable parameters. We therefore retain the same architecture in the present work to provide a consistent basis for investigating the effect of different parameter-space representations and patch geometries.
\begin{table*}[htbp]
  \centering
  \caption{CNN architecture used in this work. \textbf{C} denotes convolutional layers and \textbf{H} denotes fully connected layers. Max-pooling is applied after each convolutional layer, while dropout with a rate of 0.5 is applied to layers H5 and H6. The number of neurons in the output layer H7 equals the number of patch classes, $N_\mathrm{classes}$.}
  \label{tab:cnn_arch}
  \begin{tabular}{ccccccc}
    \toprule
    \textbf{Layer} & \textbf{Type} & \textbf{Neurons} & \textbf{Filter Size} & \textbf{Pool Size} & \textbf{Dropout} & \textbf{Activation} \\
    \midrule
    1 & C & 32  & $(1,16)$ & $(1,4)$ & ---  & ReLU \\
    2 & C & 64  & $(1,8)$  & $(1,4)$ & ---  & ReLU \\
    3 & C & 128 & $(1,8)$  & $(1,4)$ & ---  & ReLU \\
    4 & C & 256 & $(1,8)$  & $(1,4)$ & ---  & ReLU \\
    5 & H & 128 & ---      & ---     & 0.5  & ReLU \\
    6 & H & 64  & ---      & ---     & 0.5  & ReLU \\
    7 & H & $N_\mathrm{classes}$ & --- & --- & --- & Softmax \\
    \bottomrule
  \end{tabular}
\end{table*}

The network consists of four convolutional layers followed by three fully connected layers, terminating in a softmax output layer whose dimension corresponds to the number of patch classes being classified. The architecture is unchanged from the previous implementation; the extension in the present work lies in the parameter-space representation used to define the classification problem. Specifically, whereas our earlier studies considered a two-dimensional parameter space based on the component masses, here we construct patches in higher-dimensional intrinsic-parameter spaces for aligned-spin BBH signals. The complete network architecture is summarized in Table~\ref{tab:cnn_arch}.

The convolutional layers learn progressively higher-level representations of the strain data, beginning with localized waveform structures and evolving toward features relevant for discriminating between neighboring parameter-space patches. Max-pooling reduces the dimensionality of the intermediate representations and mitigates sensitivity to small shifts in waveform position and morphology. The resulting feature representation is subsequently mapped to the target patch labels through the fully connected layers and final softmax classifier.

Training is performed using the Adam optimizer~\cite{Kingma2017} with a learning rate of $10^{-4}$ and a batch size of 50. Networks are trained for a maximum of 50 epochs. We monitor validation loss during training and retain the model checkpoint at its minimum, which typically occurs within 10–15 epochs, beyond this point training loss continues to decrease while validation loss increases, indicating the onset of overfitting. All reported results correspond to this early-stopped checkpoint rather than the final epoch of training.

\section{Results}
\label{sec:results}

\subsection{First-Stage Classification: Noisy signal vs.\ Pure noise}
\label{subsec:stage1}

The objective of the first stage is to distinguish gravitational-wave signals from noise before performing template patch identification. Unlike many previous CNN-based detection studies, which generate large training sets using uniformly sampled binary parameters, the proposed approach uses only the discrete aligned-spin template bank to generate signals during training.

To evaluate the generalization capability of the classifier, testing was performed on an independent dataset comprising $5\times 10^3$ uniformly sampled binary black hole systems and $5\times 10^3$ pure-noise realizations. Consequently, the testing signals represent previously unseen waveform realizations.

\begin{figure}
\centering
\includegraphics[width=\linewidth]{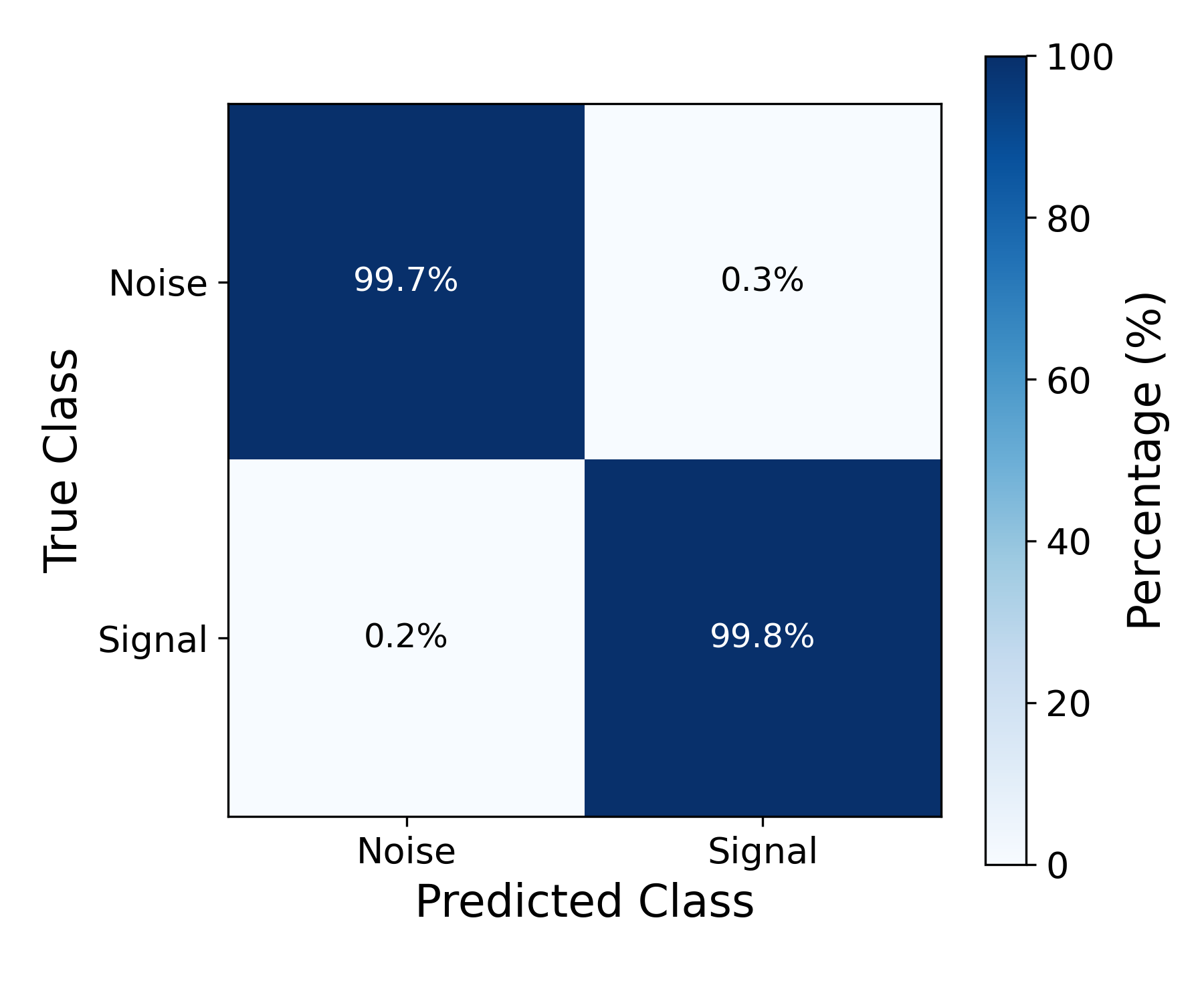}
\caption{Confusion matrix for the first-stage CNN classifier distinguishing gravitational-wave signals from detector noise. The classifier was trained using signal waveforms generated only from the aligned-spin template bank, while evaluation was performed on an independent test set consisting of uniformly sampled binary black hole systems that were not part of the template bank. The classifier achieves class-wise accuracies of 99.9\% for noise and 99.8\% for signal, demonstrating excellent generalization to unseen physical parameters.}
\label{fig:binary_class}
\end{figure}

Figure \ref{fig:binary_class} presents the confusion matrix obtained for the first-stage classifier. The classifier achieves class-wise accuracies of 99.9\% for the noise class and 99.8\% for the signal class, with only a negligible number of false positives and false negatives. These results indicate that the template-bank-based training set is sufficient for learning the distinguishing characteristics of gravitational-wave signals in Gaussian noise and generalizes effectively to uniformly sampled binary parameters.

The high detection performance demonstrates that the computationally expensive generation of large uniformly sampled training datasets is not essential for the signal-versus-noise classification task. A relatively small, physically motivated template bank can adequately represent the waveform manifold while maintaining excellent detection performance on unseen signals.

Having established that the first-stage classifier provides near-perfect discrimination between gravitational-wave signals and detector noise, we now focus on the second-stage task of identifying the appropriate template patch for the detected signal.

\subsection{Second-Stage Classification: Patch Identification}
\label{subsec:stage2}

\subsubsection{1D Patches in Chirp Mass}

\begin{figure*}[t]
\centering

\begin{minipage}[t]{0.48\textwidth}
    \centering
    \includegraphics[width=\linewidth]{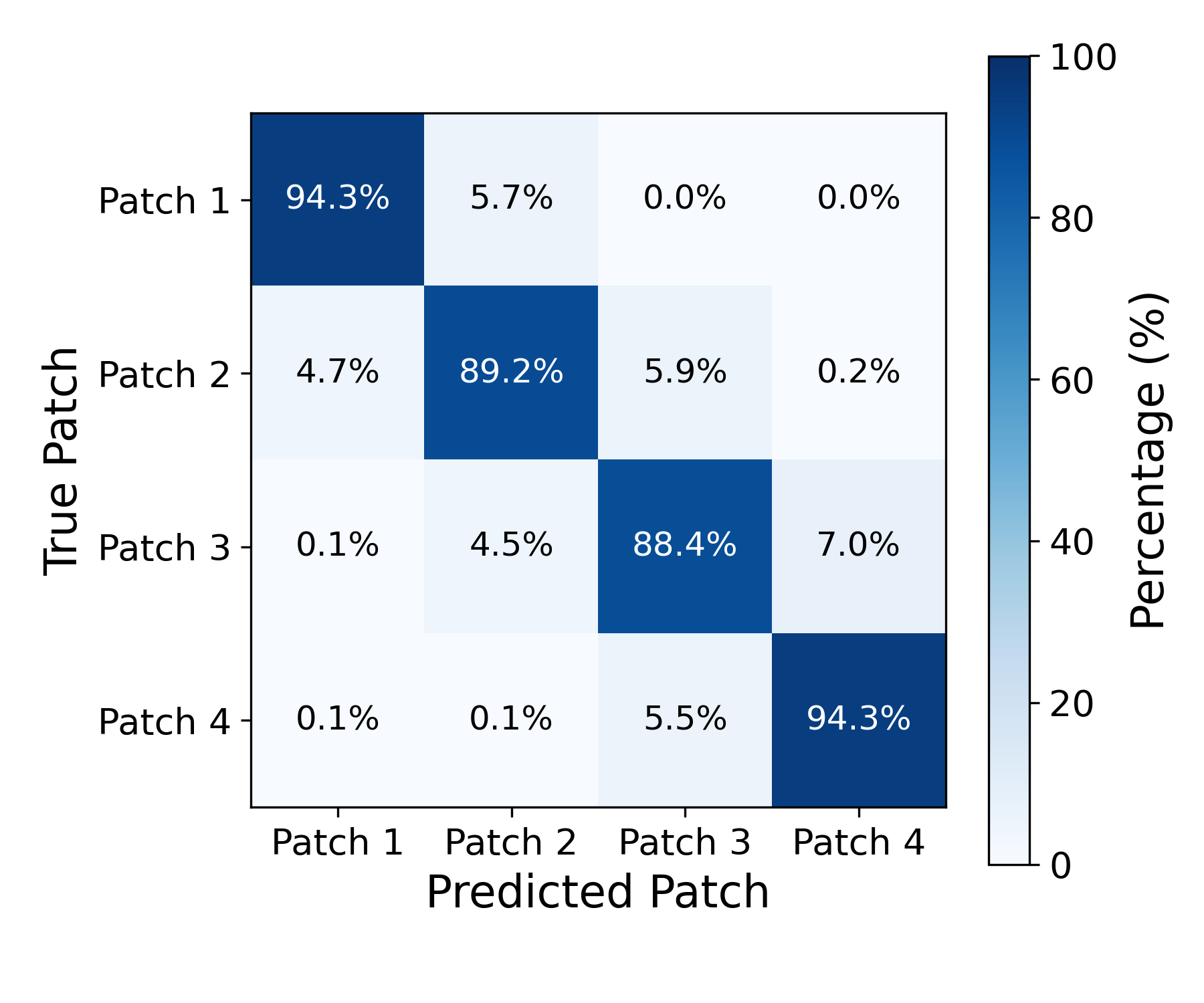}
\end{minipage}
\hfill
\begin{minipage}[t]{0.48\textwidth}
    \centering
    \includegraphics[width=\linewidth]{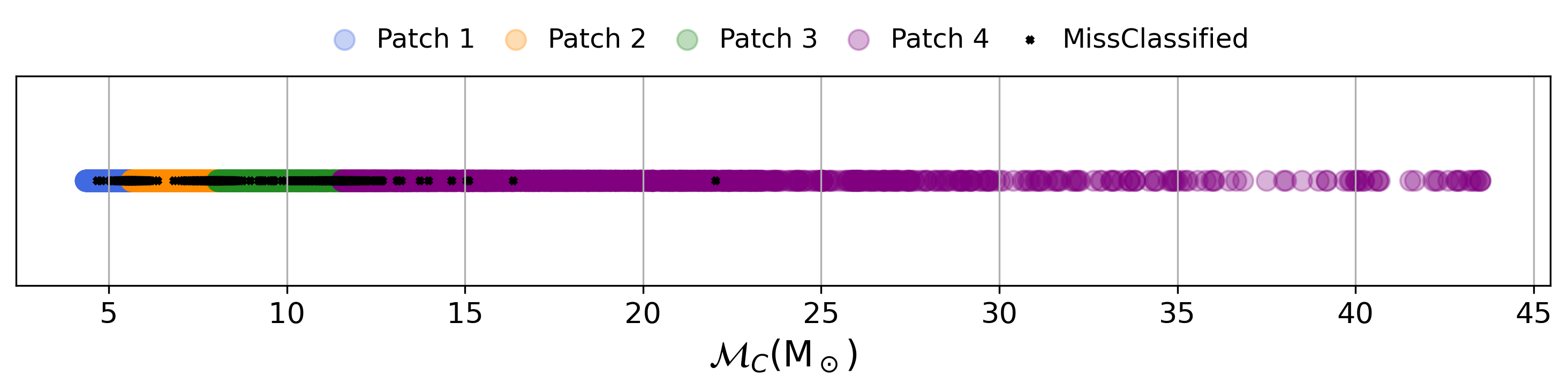}
\end{minipage}

\caption{Performance of the CNN for patch identification in the one-dimensional chirp-mass ($\mathcal{M}_c$) parameterization. The \textbf{left panel} shows the normalized confusion matrix, with patch-wise classification accuracies of 94.6\%, 89.2\%, 88.4\%, and 94.3\% for the four chirp-mass patches. The \textbf{right panel} shows the distribution of test waveforms in the chirp-mass parameter space, where colored points represent the four PCA--quantile patches and black crosses denote misclassified waveforms. The concentration of misclassified waveforms near the boundaries separating adjacent patches indicates that boundary ambiguity is the primary source of classification error, whereas the outer chirp-mass patches exhibit higher classification accuracy due to their more distinctive waveform characteristics.}

\label{fig:mc_results}
\end{figure*}

We first investigate patch construction using the chirp mass, $\Mc$, which governs the leading-order phase evolution of compact binary inspirals and is therefore expected to capture the dominant waveform morphology. Four approximately balanced patches are constructed using the PCA--quantile procedure described in Sec.~\ref{subsec:patch_schemes}. The CNN achieves patch-wise accuracies of 94.3\%, 89.2\%, 88.4\%, and 94.3\%, corresponding to an average patch-identification accuracy of approximately 91\%. The corresponding confusion matrix(left) and the distribution of misclassified samples(right) are shown in Figure ~\ref{fig:mc_results}. The confusion matrix exhibits strong diagonal dominance for the outer patches, whereas the two central patches show increased mutual confusion. This trend is also evident in the misclassification overlay, where most incorrectly classified signals are concentrated near the interfaces between adjacent chirp-mass patches. The results indicate that although chirp mass alone captures much of the information required for waveform classification, neighboring waveform families become progressively more difficult to distinguish near the intermediate patch boundaries because of the continuous dependence of the gravitational-wave signal on the intrinsic binary parameters.

\subsubsection{2D Patches}

\begin{figure}[t]
\centering

    \includegraphics[width=\linewidth]{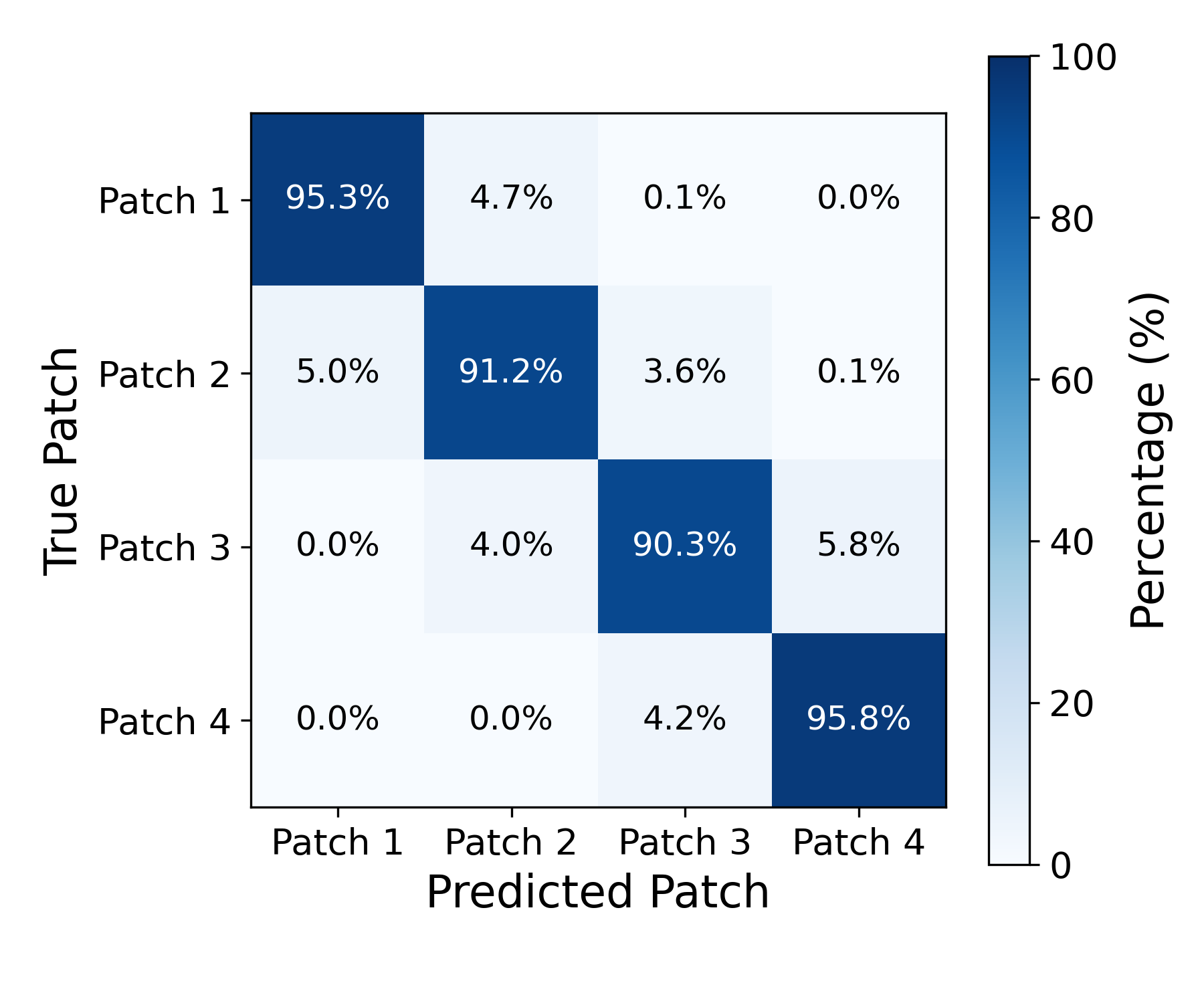}

\caption{Performance of the CNN for patch identification in the two-dimensional ($\mathcal{M}_c$, duration) parameter space. The figure shows the normalized confusion matrix, with patch-wise classification accuracies of 95.3\%, 91.3\%, 90.3\%, and 95.8\% for the four patches. The strong diagonal dominance indicates that the combined chirp-mass--duration parameterization provides a well-balanced partitioning of the parameter space and enables robust patch identification. The concentration of classification errors near the interfaces between adjacent patches indicates that the dominant source of error is boundary ambiguity rather than an inability of the CNN to learn the underlying waveform morphology. This behavior is consistent with the smooth variation of gravitational-wave signals across the parameter space and explains the high overall classification performance achieved for this parameterization.}

\label{fig:mcdur_results}
\end{figure}

\begin{figure*}
\centering
\includegraphics[width=\linewidth]{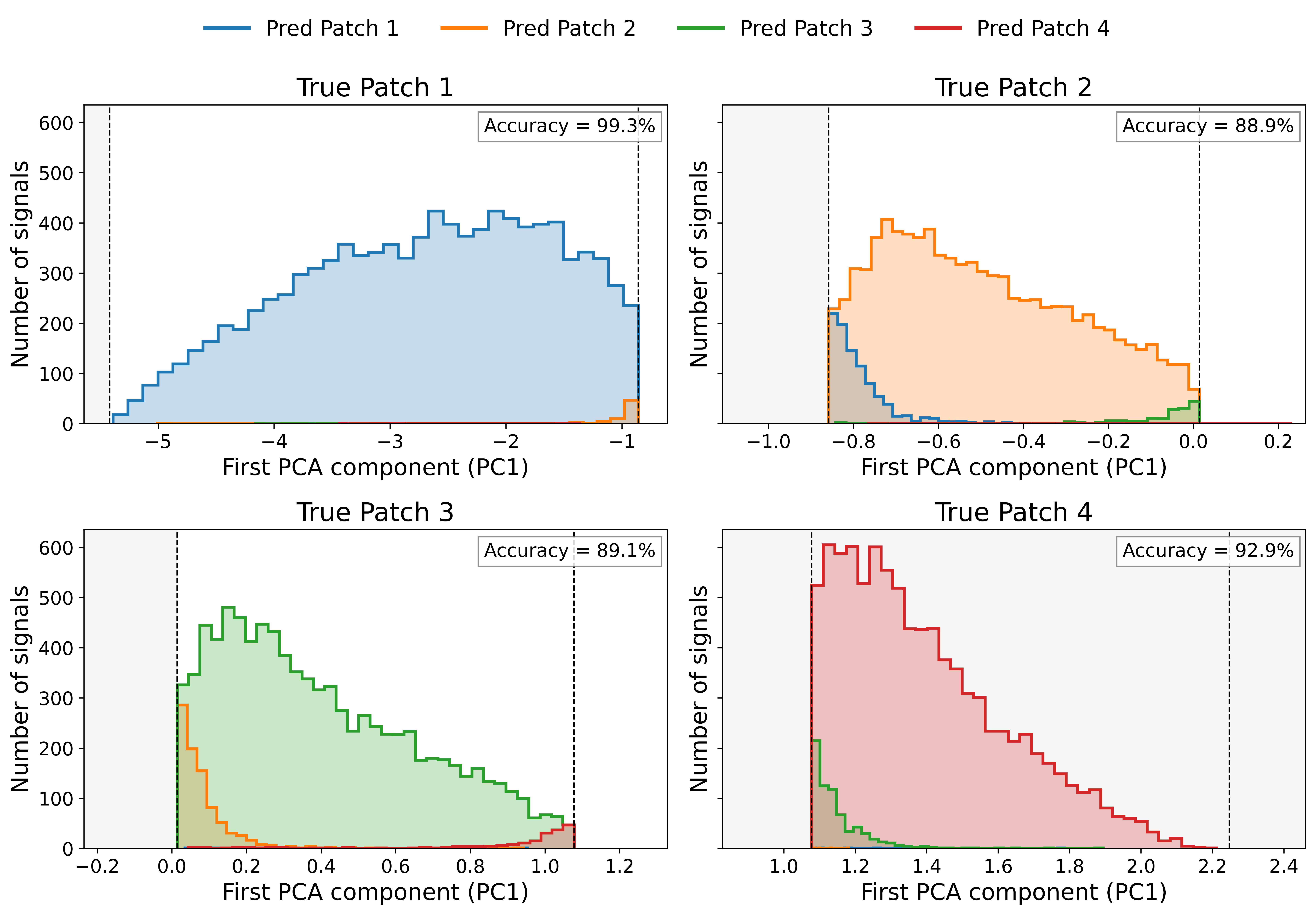}
\caption{Distribution of the test signals projected onto the first principal component (PC1) for each true template patch. The two-dimensional feature space comprising chirp mass and waveform duration was standardized and projected onto PC1 using Principal Component Analysis (PCA). The PC1 space was subsequently partitioned into four contiguous regions by three decision boundaries (vertical dashed lines), which define the template patches used for CNN training. Each subplot shows the distribution of signals from one true patch, while the colored histograms indicate the corresponding CNN-predicted patch. The boxed values denote the patch-wise classification accuracy. The figure illustrates that the CNN predictions are predominantly confined to the correct PC1 region, with the remaining misclassifications occurring primarily in neighboring patches.}

\label{fig:patch_distribution}
\end{figure*}

\begin{figure*}[t]
\centering

\begin{minipage}[t]{0.48\textwidth}
    \centering
    \includegraphics[width=\linewidth]{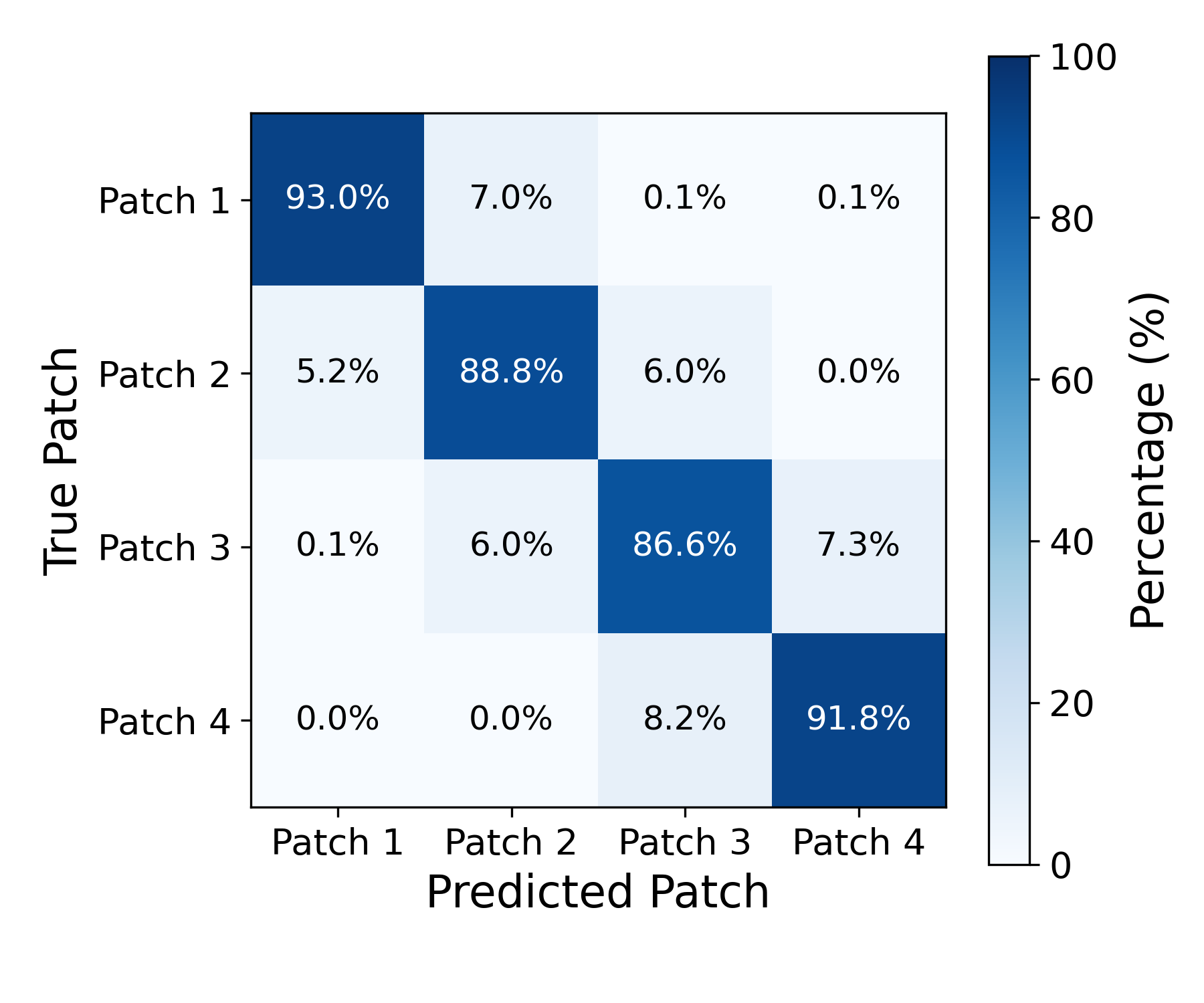}
\end{minipage}
\hfill
\begin{minipage}[t]{0.48\textwidth}
    \centering
    \includegraphics[width=\linewidth]{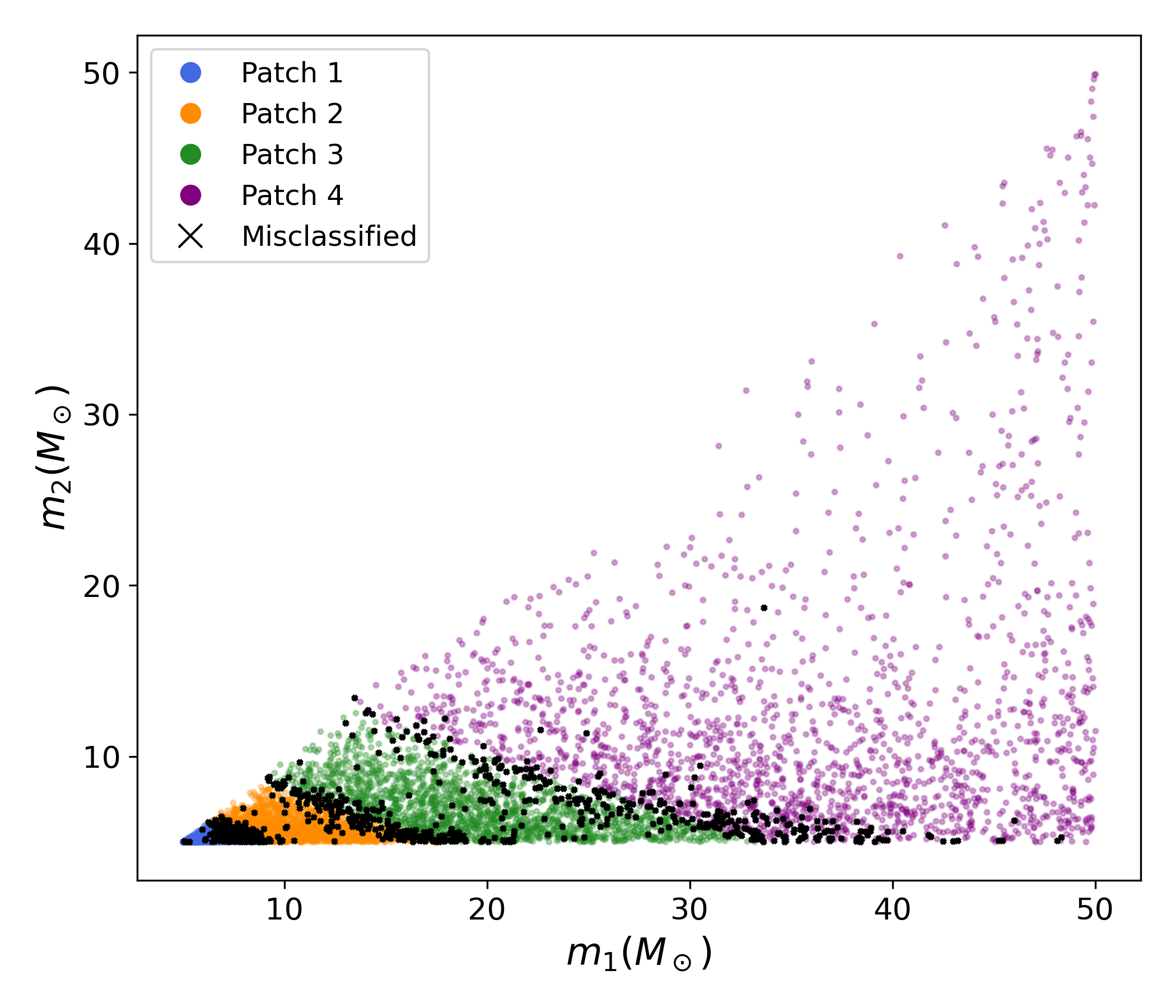}
\end{minipage}

\caption{Performance of the CNN for patch identification in the component-mass ($m_1,m_2$) parameter space. The \textbf{left panel} shows the normalized confusion matrix, with patch-wise classification accuracies of 93.0\%, 88.8\%, 86.6\%, and 91.9\% for the four patches. Although the classifier achieves high overall accuracy, the off-diagonal elements indicate a greater degree of confusion between neighboring patches than observed for the ($\mathcal{M}_c$, duration) parameterization. The \textbf{right panel} shows the distribution of test waveforms in the component-mass plane, where colored points represent the four PCA--quantile patches and black crosses denote misclassified waveforms. The concentration of misclassified samples near the interfaces between adjacent patches indicates that waveform similarity across neighboring regions of the parameter space is the primary source of classification errors. This result suggests that direct partitioning of the component-mass plane is less well aligned with the dominant directions of waveform variation learned by the CNN than parameterizations based on chirp mass and waveform duration.}

\label{fig:m1m2_results}
\end{figure*}

\begin{figure*}[t]
\centering

\begin{minipage}[t]{0.48\textwidth}
    \centering
    \includegraphics[width=\linewidth]{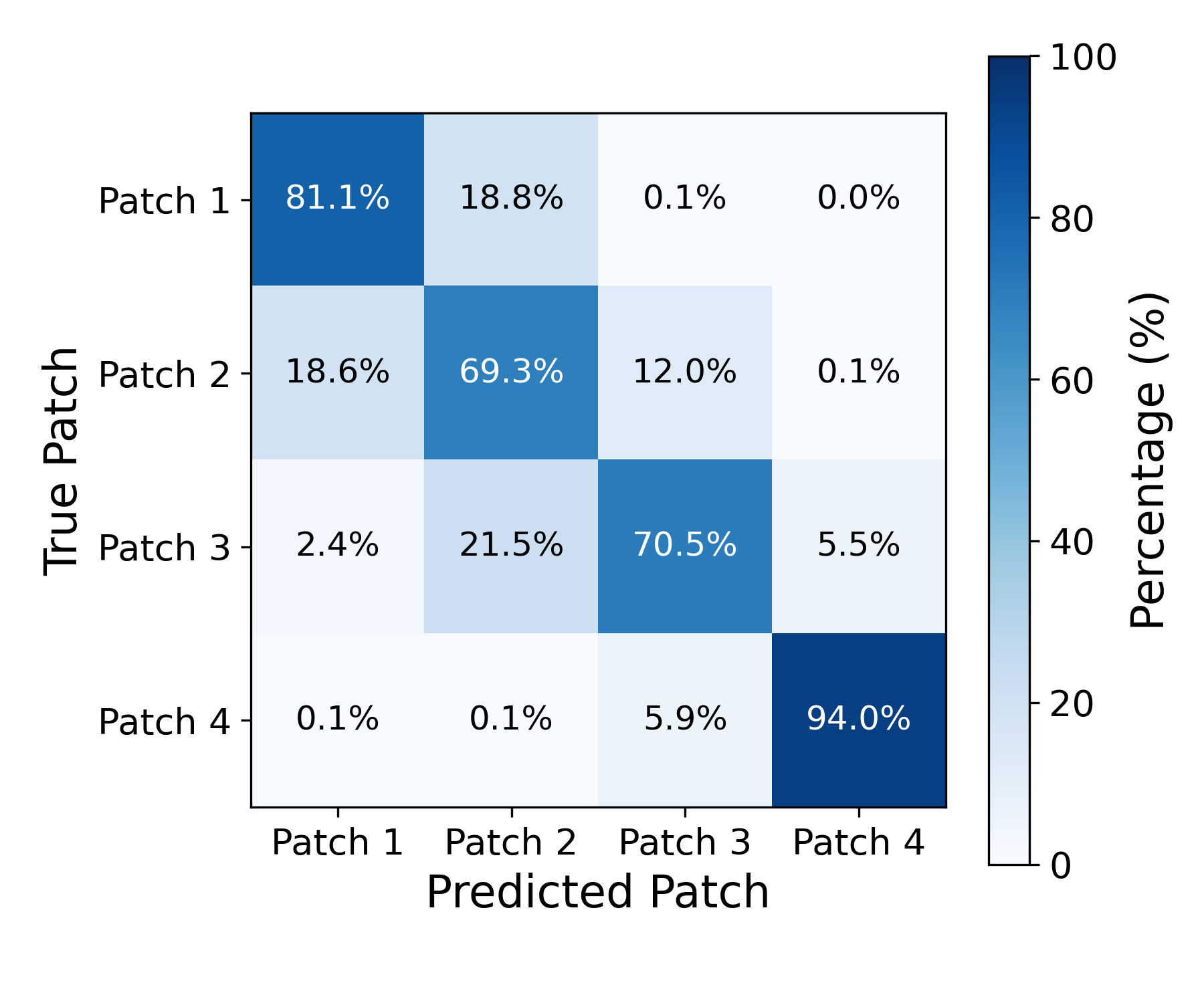}
\end{minipage}
\hfill
\begin{minipage}[t]{0.48\textwidth}
    \centering
    \includegraphics[width=\linewidth]{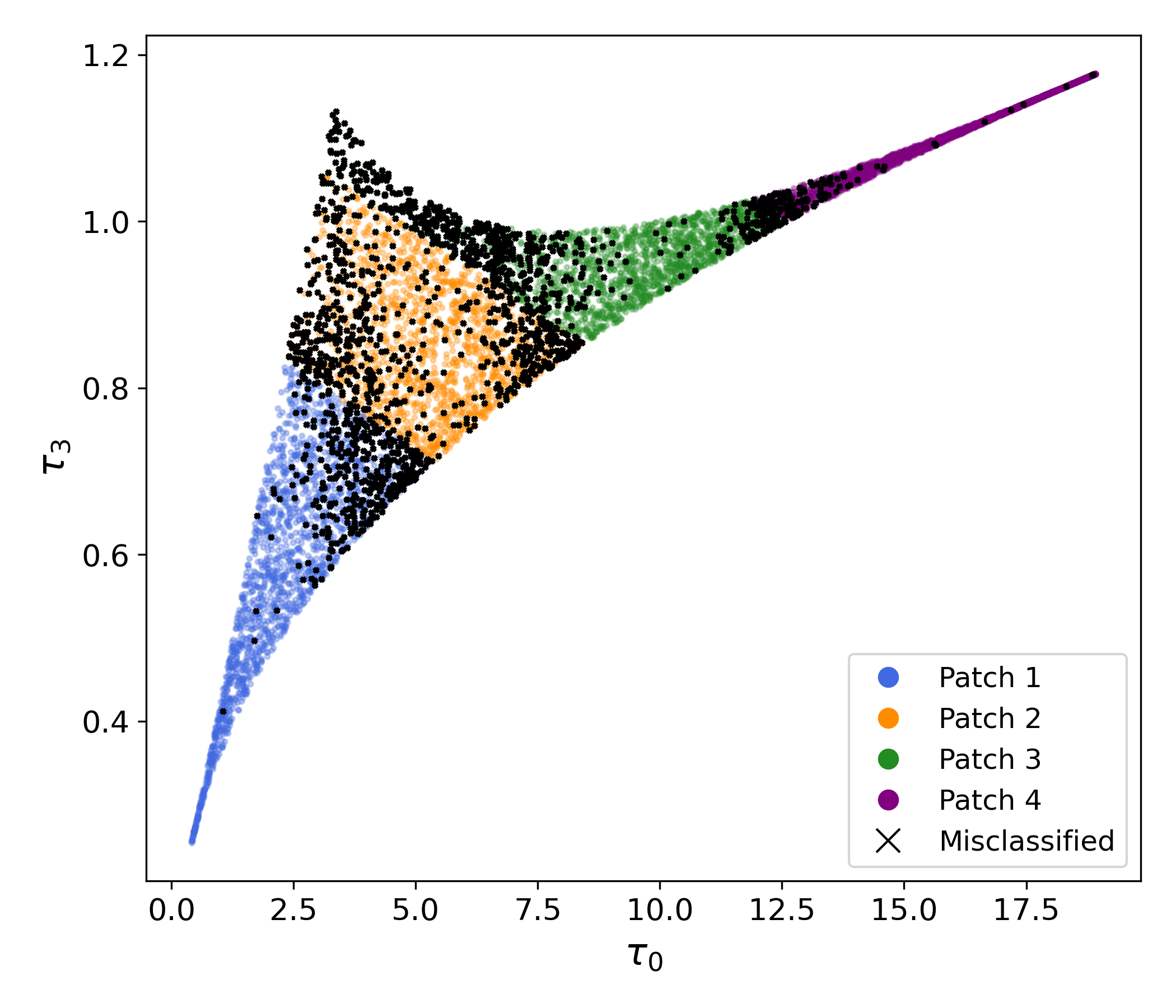}
\end{minipage}

\caption{Performance of the CNN for patch identification in the post-Newtonian chirp-time parameter space ($\tau_0,\tau_3$). The \textbf{left panel} shows the normalized confusion matrix, with patch-wise classification accuracies of 81.3\%, 69.3\%, 70.5\%, and 94.0\% for the four patches. While the outer patches are classified with high accuracy, the significantly lower accuracies of the two central patches indicate increased ambiguity between neighboring waveform classes. The \textbf{right panel} shows the distribution of test waveforms in the ($\tau_0,\tau_3$) parameter space, where colored points represent the four PCA--quantile patches and black crosses denote misclassified waveforms. The concentration of misclassified samples along the extended interfaces separating the central patches demonstrates that waveform similarity persists over larger regions of the chirp-time parameter space than in the ($\mathcal{M}_c$, duration) representation. These results suggest that, although the chirp-time coordinates provide an approximately flat metric for template-bank construction, they yield less separable waveform classes for CNN-based patch identification than parameterizations based on chirp mass and waveform duration.}

\label{fig:tau_results}
\end{figure*}

We next investigate three two-dimensional parameterizations: $(\Mc,\tau)$, $(m_1,m_2)$ and $(\tau_0,\tau_3)$. In each case, four approximately balanced patches are constructed using the PCA--quantile procedure described in Sec.~\ref{subsec:pca}. The corresponding confusion matrices(left) and misclassification distributions(right) are shown in Figs.~\ref{fig:mcdur_results}--\ref{fig:tau_results}.

\medskip
\noindent\textbf{$(\Mc, \tau)$ patches:}

Among all parameterizations investigated in this work, the $(\Mc, \tau)$ representation provides the highest classification performance. The CNN achieves patch-wise accuracies of 95.3\%, 91.3\%, 90.3\%, and 95.8\%, corresponding to an average patch-identification accuracy of approximately 93\%.

The confusion matrix(left) Fig.~\ref{fig:mcdur_results} exhibits strong diagonal dominance for all four patches, indicating that the classifier successfully distinguishes the waveform families associated with different regions of the parameter space. The corresponding misclassification distribution(right) Figure ~\ref{fig:mcdur_results} shows that incorrectly classified events are concentrated almost exclusively near the patch boundaries. This behavior suggests that the dominant source of error is the gradual transition of waveform morphology across neighboring patches rather than a limitation of the CNN itself.

Although the confusion matrix summarizes the overall classification performance, it does not reveal where the misclassified signals lie within the PCA-partitioned feature space. To investigate the spatial distribution of CNN predictions, Figure ~\ref{fig:patch_distribution} shows the distributions of the projected test signals along the first principal component for each true patch together with the predicted patch labels.

The superior performance of the $(\Mc,\tau)$ representation is physically well motivated. While the chirp mass governs the leading-order inspiral phase evolution, the waveform duration provides complementary information by encoding the combined influence of the binary masses and aligned-spin effects on the inspiral evolution. Together, these two quantities provide a compact yet highly discriminative description of waveform morphology.

\medskip
\noindent\textbf{$(m_{1}, m_{2})$ patches:}
Direct partitioning in the component-mass plane yields patch-wise accuracies of 93.0\%, 88.8\%, 86.6\%, and 91.9\%, corresponding to an average accuracy of approximately 90\%. As shown in Fig.~\ref{fig:m1m2_results}, most misclassified events occur close to the interfaces separating neighboring patches. Although the overall performance remains high, it is systematically lower than that obtained in the $(\Mc, \tau)$ representation. This result indicates that direct geometric partitioning in the component-mass plane does not align as closely with the dominant directions of waveform variation as the chirp-mass--duration representation.

\medskip
\noindent\textbf{$(\tau_{0}, \tau_{3})$ patches:}
Finally, we investigate patch construction in the post-Newtonian chirp-time coordinates $(\tau_{0}, \tau_{3})$, which are widely employed for template-bank placement owing to their approximately flat metric. The corresponding CNN achieves patch-wise accuracies of 81.1\%, 69.4\%, 70.6\%, and 94.0\%, yielding an average patch-identification accuracy of approximately 79\%.

The confusion matrix(left) in Figure ~\ref{fig:tau_results} reveals substantial confusion between the two central patches, while the misclassification overlay(right) in Figure ~\ref{fig:tau_results} shows extended regions of overlap near the corresponding patch boundaries. Although the $(\tau_{0}, \tau_{3})$ parameterization is highly effective for template-bank construction, the present results demonstrate that it provides significantly poorer waveform separability for CNN-based patch identification than the $(\Mc, \tau)$ representation. This observation suggests that parameter spaces designed to achieve efficient template placement are not necessarily optimal for machine-learning-based hierarchical searches.

\subsubsection{3D Patches in $\theta_0$--$\theta_3$--$\theta_{3s}$ Space}
Finally, we investigate patch construction in the three-dimensional parameter space $(\theta_{0}, \theta_{3}, \theta_{3s})$, which extends the post-Newtonian chirp-time coordinates by explicitly incorporating the dominant aligned-spin contribution through $\theta_{3s}$. Since these coordinates are closely related to aligned-spin template-bank construction, they provide a natural candidate for spin-dependent patch identification.

Four approximately balanced patches are constructed from the aligned-spin template-bank points using the PCA--quantile procedure described in Sec.~\ref{subsec:pca}. The resulting CNN achieves patch-wise accuracies of 77.5\%, 57.6\%, 50.8\%, and 81.2\%, corresponding to an average patch-identification accuracy of approximately 67\%.

\begin{figure}
    \centering
    \includegraphics[width=\linewidth]{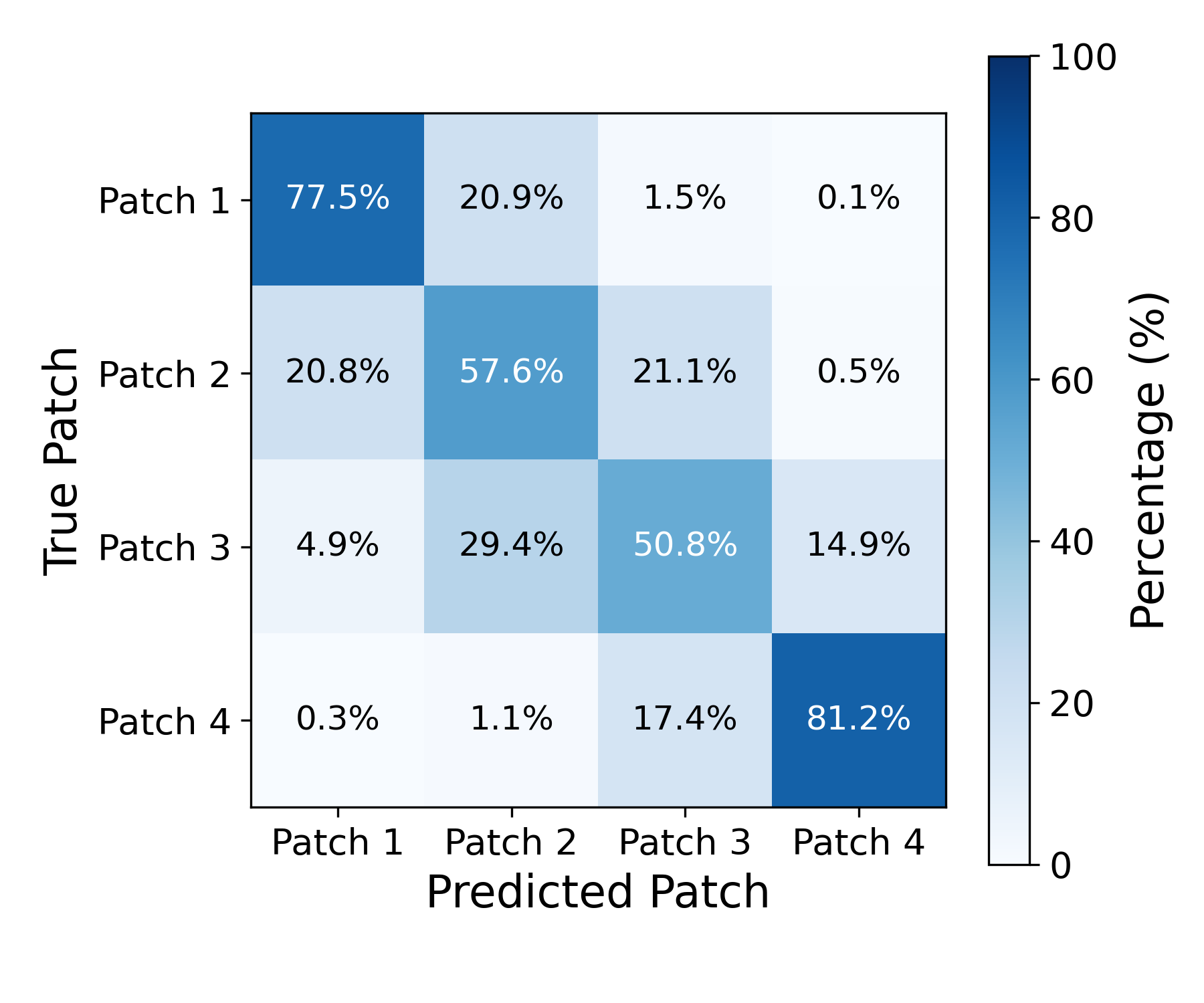}
    \caption{Confusion matrix for CNN-based patch identification in the three-dimensional parameter space $(\theta_{0},\theta_{3},\theta_{3s})$. Matrix elements are normalized by the total number of samples in each true patch and expressed as percentages. The classifier achieves patch-wise accuracies of 77.5\%, 57.6\%, 50.8\%, and 81.2\% for the four patches, corresponding to an average accuracy of approximately 67\%. Significant confusion is observed between the two central patches, while the outer patches are identified more reliably. Although the $(\theta_{0},\theta_{3},\theta_{3s})$ coordinates provide a physically motivated representation for aligned-spin template-bank construction, the corresponding patch partitioning exhibits substantially poorer waveform separability than the lower-dimensional parameterizations investigated in this work. This result suggests that coordinate systems designed for efficient template placement are not necessarily optimal for CNN-based patch identification.
}

    \label{fig:theta_confusion}
\end{figure}

Figure~\ref{fig:theta_confusion} presents the corresponding confusion matrix. Significant confusion is observed between the two central patches, whereas the outer patches are identified with comparatively higher accuracy. Compared with all lower-dimensional parameterizations investigated in this work, the $(\theta_{0}, \theta_{3}, \theta_{3s})$ representation yields the lowest classification performance.

Although the $(\theta_{0}, \theta_{3}, \theta_{3s})$ coordinates are physically well motivated for aligned-spin template-bank construction, the present results indicate that the corresponding PCA--quantile partitioning does not provide the same degree of waveform separability as the lower-dimensional parameterizations. This observation suggests that coordinate systems designed to produce efficient template placement are not necessarily optimal for CNN-based hierarchical patch identification.

Table~\ref{tab:accuracy_summary} summarizes the average classification accuracy obtained for all parameter-space representations investigated in this work.

\begin{table}[htbp]
\centering
\caption{Summary of the average patch-identification accuracy obtained for the different parameter-space representations investigated in this work. In all cases, the parameter space is partitioned into four approximately balanced patches using the PCA--quantile procedure described in \ref{subsec:pca}}
\label{tab:accuracy_summary}
\begin{tabular}{lcc}
\toprule
\textbf{Patch Scheme} & \textbf{Dimension} & \textbf{Average Accuracy (\%)}\\
\midrule
Chirp mass $\Mc$ & 1D & 91.55 \\
$\Mc$--Duration & 2D & 93.15 \\
$m_1$--$m_2$ & 2D & 90.05 \\
$\tau_0$--$\tau_3$ & 2D & 78.72 \\
$\theta_0$--$\theta_3$--$\theta_{3s}$ & 3D & 66.77 \\
\bottomrule
\end{tabular}
\end{table}

The results presented above demonstrate that the choice of parameter-space representation significantly affects the accuracy of CNN-based patch identification. Among the representations investigated, the $(\Mc,\tau)$ parameterization consistently provides the highest classification accuracy, followed by the $(m_{1}, m_{2})$ and $\Mc$ representations. In contrast, the post-Newtonian coordinate systems $(\tau_{0}, \tau_{3})$ and $(\theta_{0}, \theta_{3}, \theta_{3s})$, although well suited for template-bank construction, exhibit substantially poorer waveform separability for the present machine-learning framework. These findings suggest that parameterizations optimized for matched-filter template placement are not necessarily the most effective choices for the CNN-assisted hierarchical gravitational-wave search framework. 

\section{Discussion}
\label{sec:discussion}
\subsection{Parameter-Space Representation and Waveform Separability}
The principal finding of this work is that (a) the sparse and relatively small template bank can be used to train a CNN model for a binary classification task between pure noise and a noisy signal, and  (b) the choice of parameter-space representation significantly influences the accuracy of the CNN-based patch identification task. Since all parameterizations are constructed from the same aligned-spin template bank and evaluated using an identical CNN architecture and training procedure, the observed differences in classification performance can be attributed primarily to the underlying coordinate representation rather than differences in the sampled parameter distribution. Although all parameterizations describe the same intrinsic aligned-spin BBH parameter space, they differ substantially in their ability to separate waveform families into distinct classes. Among the representations investigated, the $(\Mc, \tau)$ parameterization consistently yields the highest classification accuracy. This result can be understood from the physical information these variables contain. The chirp mass governs the leading-order phase evolution of the inspiral, while the waveform duration provides complementary information reflecting the combined influence of mass ratio and aligned-spin effects on the orbital evolution. Together, these quantities provide a compact representation that closely correlates with the dominant observable characteristics of the waveform, thereby improving separability between neighboring patch classes.
Although the component masses uniquely determine the intrinsic waveform, the dependence of the waveform morphology on $(m_{1}, m_{2})$ is highly nonlinear. Consequently, neighboring regions in the component-mass plane do not necessarily correspond to maximally distinct waveform families, leading to increased overlap between adjacent patches and slightly higher misclassification rates than in the $(\Mc,\tau)$ representation. Perhaps the most interesting observation concerns the post-Newtonian coordinate systems $(\tau_{0}, \tau_{3})$ and $(\theta_{0}, \theta_{3}, \theta_{3s})$. These coordinates were originally developed to facilitate template-bank construction by producing an approximately flat metric over the intrinsic parameter space. Our results demonstrate that this property does not necessarily translate into improved CNN classification performance. Although these coordinates provide an efficient representation for template placement, the corresponding PCA--quantile partitions exhibit substantially poorer waveform separability than the $(\Mc, \tau)$ and $(m_{1}, m_{2})$ representations. In particular, for the three-dimensional $(\theta_{0},\theta_{3}, \theta_{3s})$ parameterization, the resulting partitions exhibit considerable overlap in the spin-dependent direction, reducing the distinctiveness of the waveform classes presented to the CNN. These findings suggest that coordinate systems optimized for matched-filter template placement are not necessarily optimal for CNN-assisted hierarchical searches. More generally, these results indicate that waveform separability, rather than geometric regularity of the parameter space, is the key factor governing CNN-based patch identification. An effective parameterization is therefore one that groups together waveforms with similar morphology while maximizing the distinction between neighboring patches. This observation provides a useful guideline for designing parameter-space representations in future CNN-assisted gravitational-wave search pipelines.

\subsection{Implications for Hierarchical Matched-Filter Searches}
\label{subsec:cost_reduction}

The framework proposed in this work is intended to operate as a template-bank localization stage preceding, rather than replacing, the matched-filter component of a hierarchical compact-binary-coalescence (CBC) search. The CNN performs coarse localization of a gravitational-wave signal in the intrinsic-parameter space by identifying the template-bank patch most likely to contain it. The subsequent matched-filter search can then be restricted to the corresponding subset of templates. In this architecture, the CNN does not replace the matched-filter detection statistic or the downstream procedures used to assess candidate significance; rather, it determines the region of the template bank over which the conventional matched-filter calculation is performed.

In this framework for the best-performing $(\Mc, \tau)$ representation, the CNN achieves an average patch-localization accuracy of approximately $93\%$. For correctly localized signals, the subsequent matched-filter search therefore needs to examine approximately one quarter of the complete template bank, corresponding to a $75\%$ reduction in the number of templates searched. This value represents a reduction in the matched-filter template workload under the adopted equal-patch construction; it should not be interpreted as a direct measurement of the wall-clock computational cost or latency of an end-to-end CBC search, since the filtering cost of individual templates can depend on waveform duration, sampling rate, FFT implementation, batching, and other details of the search infrastructure.

An important outcome of this work is that the utility of the localization stage depends not only on its overall patch-localization accuracy but also on the spatial structure of its errors within the intrinsic-parameter space. Examination of the confusion matrices together with the patch distributions shown in Fig.~\ref{fig:patch_distribution} reveals that the majority of the localization errors occur between adjacent patches, whereas confusion between non-neighbouring patches is comparatively rare. Thus, the remaining errors are not distributed arbitrarily throughout the template bank but are predominantly associated with the common boundaries between neighbouring patches. This observation has a direct implication for the hierarchical search: a patch-localization error need not result in a missed signal if the true signal lies sufficiently close to the boundary between the predicted and true patches.

This motivates a boundary-overlap strategy in which the matched-filter search is extended by a small number of templates from neighbouring patches. To estimate the potential benefit of this strategy, we performed a post hoc analysis using the test dataset. For a given overlap percentage, the specified fraction of templates closest to each shared patch boundary was included from the neighbouring patch. Since the overlap is defined in terms of template count rather than a fixed geometric distance, the corresponding width in the PCA coordinate varies according to the local template density. For each test signal, the predicted patch, together with the prescribed neighbouring overlap regions, was then treated as the hypothetical matched-filter search region. An adjacent-patch localization error was potentially recoverable when the true signal location fell within this expanded search region. The resulting quantity is therefore an estimated fraction of test signals covered by the proposed search region under the post hoc recovery assumption; it is not an improvement in the measured CNN patch-localization accuracy.

\begin{figure}
\centering
\includegraphics[width=\linewidth]{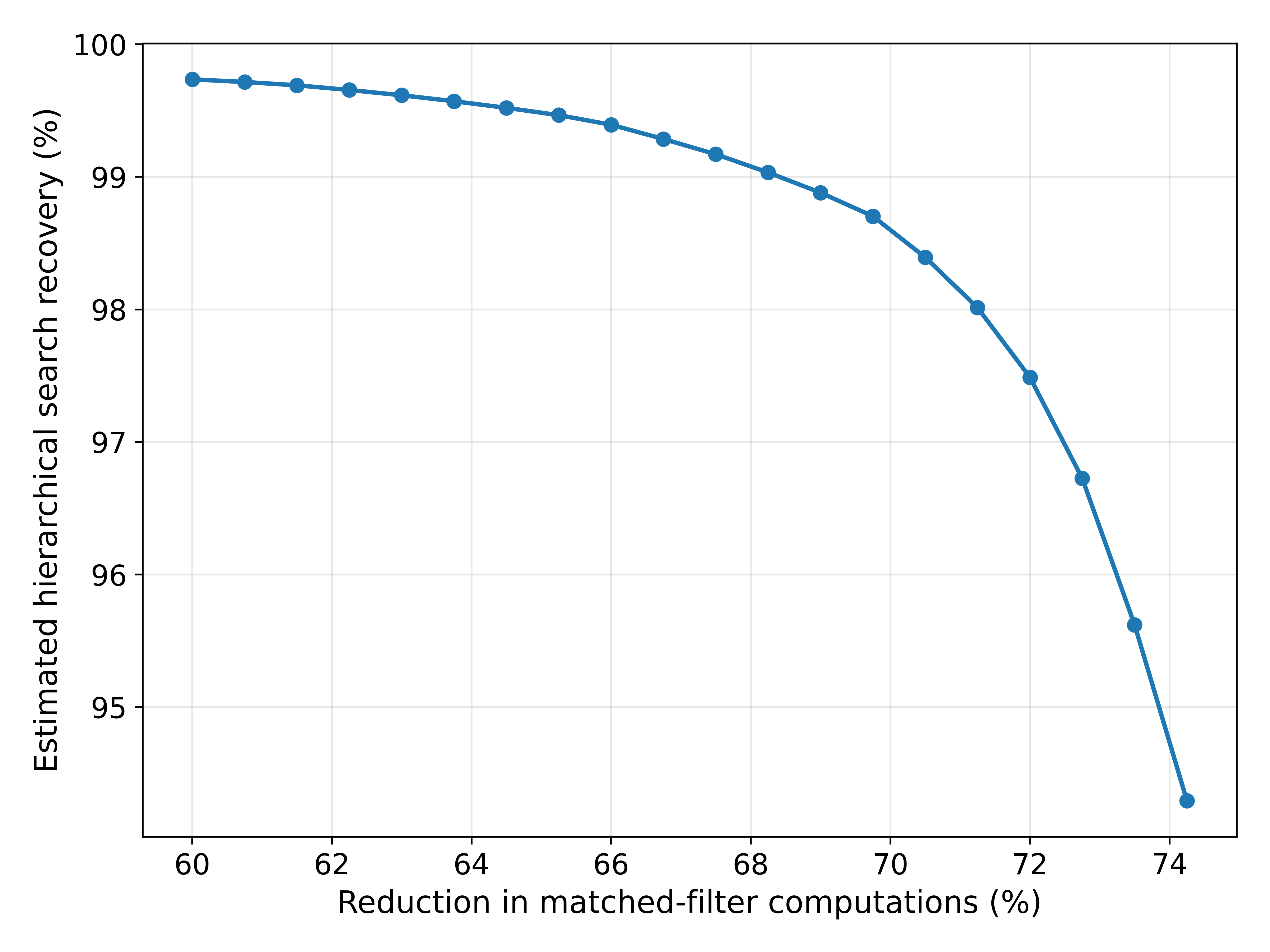}
\caption{Estimated trade-off between hierarchical search recovery and matched-filter computational reduction obtained using a boundary-overlap strategy. Following CNN-based patch prediction, matched filtering is assumed to be performed over the predicted template-bank patch together with a fixed percentage of templates closest to each shared patch boundary from the neighbouring patches. The overlap percentage is identical for all boundaries, while the corresponding geometric overlap width in the PCA coordinate varies naturally according to the local template density. The numerical values at representative operating points are listed in Table~\ref{tab:overlap}.}
\label{fig:cost_recovery}
\end{figure}

Representative operating points obtained from this analysis are summarized in Table~\ref{tab:overlap}, while the trade-off between the estimated recoverable fraction and the reduction in the number of templates searched is shown in Fig.~\ref{fig:cost_recovery}.

\begin{table}[htbp]
\centering
\caption{Representative operating points illustrating the trade-off between template overlap, estimated hierarchical search recovery, and matched-filter (MF) computation reduction for the proposed boundary-overlap strategy. All values are expressed as percentages. The template overlap denotes the fraction of neighbouring templates included from each shared patch boundary during the subsequent matched-filter stage. The estimated search recovery is obtained by combining the correctly classified signals with adjacent-patch misclassifications that are assumed to be recovered by the overlap search. The MF computation reduction is measured relative to a conventional matched-filter search over the complete template bank.}
\label{tab:overlap}
\begin{tabular}{ccc}
\toprule
\textbf{Overlap} &
\textbf{\shortstack{Estimated\\Recovery}} &
\textbf{\shortstack{MF Computation\\Reduction}} \\
\midrule
0  & 93.00 & 75.00 \\
2  & 95.62 & 73.50 \\
4  & 97.49 & 72.00 \\
\textbf{6} & \textbf{98.39} & \textbf{70.50} \\
8  & 98.88 & 69.00 \\
10 & 99.17 & 67.50 \\
\bottomrule
\end{tabular}
\end{table}

As expected, increasing the template overlap increases the estimated recoverable fraction at the expense of searching additional templates. Without overlap, the measured CNN patch-localization accuracy is approximately $93\%$, while restricting the search to one quarter of the template bank yields a $75\%$ reduction in the number of templates searched. With a $6\%$ overlap of neighbouring templates, the post hoc analysis yields an estimated recoverable fraction of approximately $98.4\%$, while retaining a reduction of approximately $70.5\%$ in the number of templates searched relative to the complete template bank. Increasing the overlap further produces progressively smaller increases in the estimated recoverable fraction while requiring a larger template-search workload.

The importance of this result is that the utility of the CNN localization stage is not determined solely by its overall localization accuracy. A CNN with imperfect patch localization can remain effective for a hierarchical search if its localization errors are spatially concentrated near the boundaries between neighbouring patches. In such a case, a relatively small extension of the subsequent matched-filter search can cover a substantial fraction of the otherwise incorrectly localized signals. The boundary-overlap strategy, therefore, provides a mechanism for trading an increased template-search workload against the estimated coverage of signals affected by patch-localization errors.

The recoverable fractions obtained here should not be interpreted as end-to-end detection efficiencies or sensitivities at a specified false-alarm rate. The present analysis does not perform the subsequent matched-filter searches over the expanded template regions and therefore does not account for the matched-filter detection statistic, background distribution, false-alarm probability, or detection threshold. In particular, changing the number of templates searched can modify the number of matched-filter trials and the associated background statistics. Therefore, establishing the effect of the proposed localization and boundary-overlap strategies on detection efficiency and search sensitivity requires an end-to-end, injection-based evaluation of the complete hierarchical search pipeline.

\section{Conclusion}
\label{sec:conclusion}

In this work, we extended our previously proposed CNN-based hierarchical gravitational-wave search framework ~\cite{Verma2022} to aligned-spin binary black hole systems and investigated how the choice of intrinsic-parameter representation affects CNN-based patch identification.
Our primary aim is not only to detect the astrophysical GW signal in the noisy data, but also to constrain the corresponding intrinsic parameter regimes associated with the signal. The constraint on the intrinsic parameter would narrow the parameter space in which an expensive matched-filter scheme would be plausible. Further, the constraint also helps reduce the prior range for the Bayesian inference, which is helpful for the first parameter estimation run and electromagnetic (EM) follow-up. 
Motivated by template bank placement across different transformations of the same intrinsic parameter space, we further investigated how a sparse template bank can be used as training data for classifying noisy signals vs. pure noise. A relatively small and sparse template bank can achieve high classification accuracy (~99$\%$), indicating that no large amount of training data is needed. Further, we investigated that different representations of the same physical parameter space can lead to substantially different localization performance of the CNN model. Our results reveal that CNN-based localization achieves the best accuracy in a parameter space that is not optimal for an efficient template-bank placement algorithm. 

It is further expected that the localization errors are predominantly confined to neighbouring template-bank patches. This spatial structure has direct implications for a hierarchical matched-filter search: the CNN need not achieve perfect patch localization if errors near shared patch boundaries can be accommodated by extending the subsequent matched-filter search into a small region of neighboring tiles. Our post hoc boundary-overlap analysis demonstrates the potential of this strategy to retain a substantial reduction in the number of templates searched while covering a large fraction of the signals affected by patch-localization errors. The reported recoverable fraction should, however, be regarded as a test-set-based estimate rather than an end-to-end detection efficiency.

The results, therefore, support a hierarchical search architecture in which the CNN acts as a coarse template-bank localization stage preceding conventional matched filtering. The matched-filter statistic and downstream candidate-ranking and background-estimation procedures can remain unchanged in principle, while the CNN reduces the region of intrinsic-parameter space requiring detailed matched-filter evaluation. This provides a complementary role for machine learning in CBC searches: rather than replacing matched filtering, it can reduce the template-search workload before the conventional detection stage.

Before such a framework can be incorporated into an operational GW search, its performance must be established in an end-to-end pipeline. In particular, future work must quantify the actual computational latency and resource reduction achieved by template pre-selection, evaluate the effect of boundary overlap on the matched-filter background and false-alarm statistics, and determine the resulting detection efficiency and sensitivity through injection-based studies. Validation with non-Gaussian detector noise, multi-detector data, and established CBC search infrastructures, such as PyCBC ~\cite{alex} and GstLAL or SGNL ~\cite{GstLAL, SGNL}, will be essential. Such studies will determine whether CNN-based localization can provide a practical front end for computationally efficient, low-latency hierarchical GW searches.

\section*{Acknowledgements}
The authors' CV and GG acknowledge the use of computational resources provided by the High Performance Computing (HPC) facility at the Inter-University Centre for Astronomy and Astrophysics (IUCAA), Pune, India. We also thank the IUCAA HPC support team for their valuable assistance.

\bibliography{references}

\end{document}